\documentclass[twocolumn,groupedaddress,amsmath,aps,longbibliography]{revtex4-1}
\usepackage{amsmath,amsfonts}
\usepackage{dcolumn}
\usepackage{bm}
\usepackage{color}
\usepackage{multirow}

\usepackage{graphicx}
\usepackage{hyperref}
\usepackage{float} 
\usepackage[utf8x]{inputenc}

\begin{document}


\title{Electron-phonon superconductivity in CaBi$_2$ and the role of spin-orbit interaction.}



\author{Sylwia Goł{\c{a}}b}
\author{Bartłomiej Wiendlocha}
\email{wiendlocha@fis.agh.edu.pl}
\affiliation{Faculty of Physics and Applied Computer Science,
AGH University of Science and Technology, al. Mickiewicza 30, 30-059 Krakow, Poland}

\date{\today}

\begin{abstract}
CaBi$_2$ is a recently discovered type-I superconductor with $T_c=2$~K and a 
layered crystal structure. In this work electronic structure, lattice dynamics 
and electron-phonon interaction are studied, with a special attention paid to the influence 
of the spin-orbit coupling (SOC) on above-mentioned quantities.
We find, that in the scalar-relativistic case (without SOC), electronic structure and electron-phonon interaction 
show the quasi-two dimensional character. 
Strong Fermi surface nesting is present, which leads to appearance 
of the Kohn anomaly in the phonon spectrum and enhanced electron-phonon coupling for the phonons propagating in the Ca-Bi atomic layers.
However, strong spin-orbit coupling in this material changes the topology of the Fermi surface, reduces the nesting
and the electron-phonon coupling becomes weaker and more isotropic.
The electron-phonon coupling parameter $\lambda$ is 
reduced by SOC almost twice, from 0.94 to 0.54, giving even stronger effect on the superconducting critical 
temperature $T_c$, which drops from 5.2~K (without SOC) to 1.3~K (with SOC). 
Relativistic values of $\lambda$ and $T_c$ remain in a good agreement with experimental findings, 
confirming the general need for including SOC in analysis of the electron-phonon interaction in materials containing heavy elements.
\end{abstract}


\maketitle

\section{Introduction}

Elemental bismuth has unusual electronic properties. It is a semimetal, crystallizing in a diatomic, rhombohedral structure, which is a result 
of a Peierls-Jones distortion~\cite{Bi-jones}. Bi exhibits the strongest diamagnetism of all elements in the normal state (susceptibility 
$\chi\sim10^{-5}$ emu) related to the large spin-orbit coupling effects \cite{Bi-szescian}, as it has the highest atomic number ($Z = 83$) 
of all non-radioactive elements.
In its band structure one can find 
Dirac-like electronic states with small effective mass \cite{Bi-szescian} and large mobility. 
Bismuth has very low charge carrier density of electrons and holes (about $10^{-5}$ carier per atom) 
and its Fermi surface consists of three electronic and one hole pockets~\cite{bi-fs,bi-in}. 
As the electronic pockets lose their symmetry in the magnetic field, Bi was recently proposed as a ''valleytronic''
material, where contribution of each electronic pocket to the charge transport may be tuned by the magnetic field~\cite{Bi-valleytronic}.  
As far as the superconductivity is concerned, it was discovered long time ago, that amorphous bismuth is a superconductor with relatively high
$T_c$ = $6$~K~\cite{Buckel1954,Bi-amorphous-sc}.
On the other hand, crystalline bismuth was long considered not to be a superconductor, although finally it was found, that 
superconductivity occurs in ultra-low temperatures, below $T_c$ = $0.53$~mK~\cite{Bi-sc}. 

There are many bismuth-based high-temperature superconductors, like 
Bi$_2$Sr$_2$CaCu$_2$O$_8$, where Bi$_2$O$_2$ layer plays a role of a charge reservoir~\cite{bi-high-tc-sc}. 
Among the low-temperatures superconductors, we find several Bi-based families, including  
$A$Bi$_3$, with $A$ = Sr, Ba, Ca, Ni, Co, La~\cite{abi3,abi3-exp-1st,nibi3,labi3,cobi3},
$A$Bi with $A$ = Li, Na  \cite{libi,nabi}, or $A$Bi$_2$ with $A$ = K, Rb, Cs, and Ca.
In the last family, KBi$_2$, RbBi$_2$, and CsBi$_2$, with $T_c$ = 3.6~K, 4.25~K and $4.75$~K 
respectively, adopt cubic $fcc$ structure~\cite{abi2-tc}, while our title compound CaBi$_2$, with $T_c = 2.0$~K, is orthorhombic~\cite{cabi2}.

In recent years Bi compounds have attracted much attention as candidates for topological 
materials or topological superconductors.  Among them we may find the well-known examples of semiconducting
Bi$_{1-x}$Sb$_x$ alloy, or the "thermoelectric" tetradymites Bi$_2$Te$_3$, Bi$_2$Se$_3$ \cite{coloqium,heremans_rev} 
and their relatives, like Sr$_x$Bi$_2$Se$_3$~\cite{srbise}.
Also $A_3$Bi$_2$ ($A$=Ca, Sr, Ba) compounds are considered as 3D topological insulators~\cite{m3bi2}. 
Moreover, topological states are present eg. in ThPtBi, ThPdBi and ThAuBi \cite{ThXY} (topological metals), 
HfIrBi \cite{hfirbi} (topological semimetal), Bi$_4$I$_4$ \cite{bii-1d} (quasi-1D 
topological insulator). All these examples show, that bismuth-based materials offer a variety of interesting physical properties, 
usually related to the strong relativistic effects.

In this work we focus on CaBi$_2$ compound, recently reported~\cite{cabi2} to be a type-I superconductor, with $T_c=2.0$~K. 
The key problem we would like to address is what is the effect of the spin-orbit coupling (SOC) on the electron-phonon interaction and superconductivity in this material.
In order to do so, electronic structure, phonons and the electron-phonon coupling function are computed, in both scalar-relativistic~\cite{scalar} (without SOC) and relativistic (including SOC) way, and we found, that SOC indeed has a very strong impact on the computed quantities.
In the scalar-relativistic case, electronic structure and electron-phonon interaction 
show the quasi-two dimensional character, with significantly enhanced electron-phonon coupling for the phonons propagating in the Ca-Bi atomic layers.
However, strong spin-orbit coupling in this material changes the topology of the Fermi surface, 
{indirectly}
making the electron-phonon interaction more three-dimensional and weaker, and the computed electron-phonon coupling constant $\lambda$ is reduced nearly twice.

\section{Computational details}

\begin{figure}[t]
\includegraphics[width=0.45\textwidth]{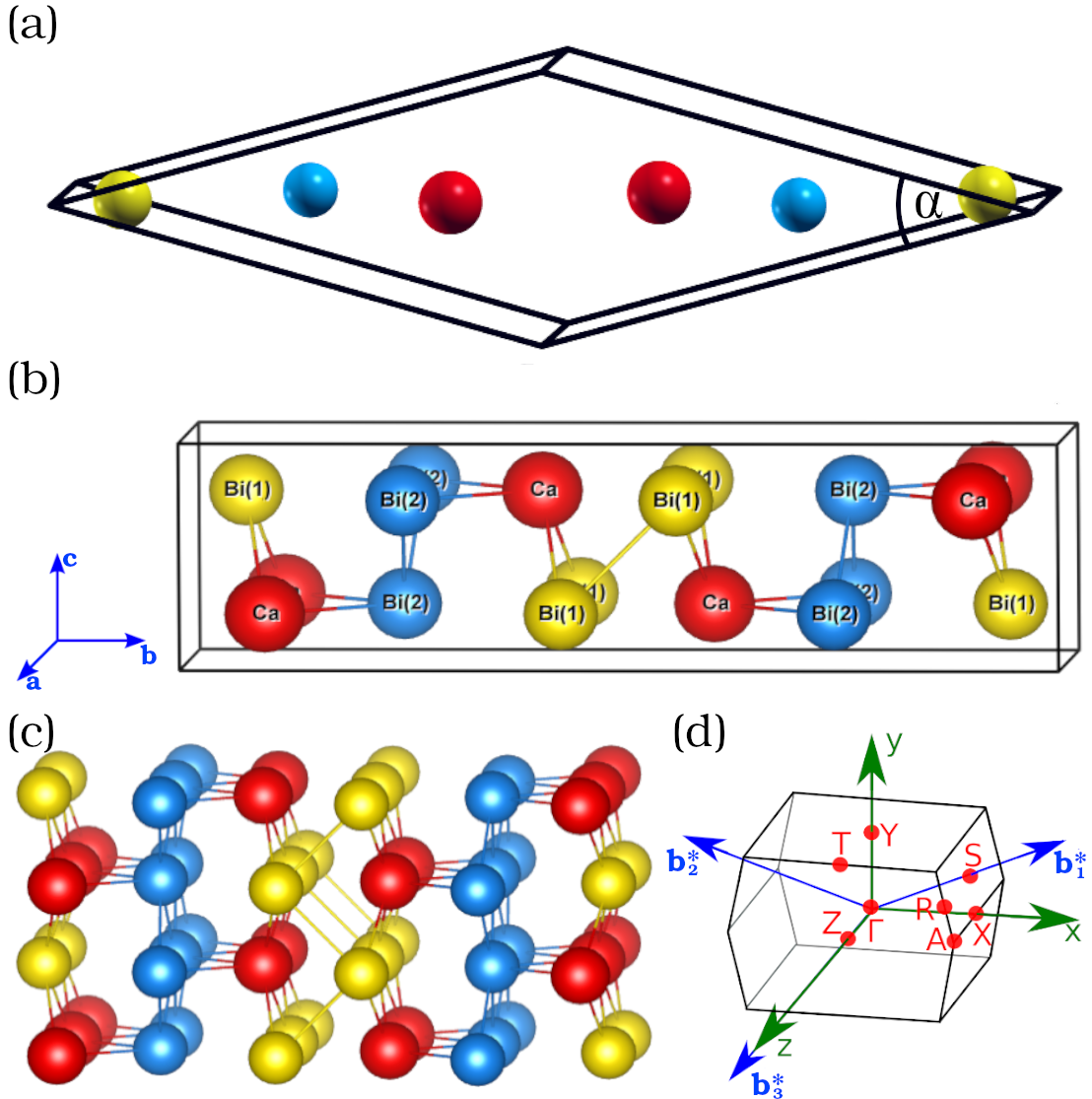}
\caption{The CaBi$_2$ crystal structure. Ca, Bi(1) and Bi(2) atoms are marked by red, 
yellow and blue balls, respectively. Panel (a) shows the primitive cell; (b) the conventional unit cell; (c) eight unit cells 
stacked to show the Ca-Bi(1) and Bi(2) atomic layers; (d) Brillouin zone of space group no. 63 with high-symmetry {\bf k}-points marked. ${\bf b^*}$ are the reciprocal primitive vectors, while $x,y,z$ are the Cartesian vectors in reciprocal space, parallel to the conventional unit cell {\bf a}, {\bf b}, {\bf c} vectors.}\label{fig:struct}
\end{figure}

CaBi$_2$ forms an orthorhombic ZrSi$_2$-type structure (space group \textit{Cmcm}, no. 63), 
which is shown in Fig.~\ref{fig:struct}. 
The primitive cell of CaBi$_2$ is shown in Fig.~\ref{fig:struct}(a) and contains 2 formula units (f.u.). There are two inequivalent positions of Bi atoms, denoted in this work as Bi(1) and Bi(2), whereas Ca atoms occupy one position. 
The base-centered conventional unit cell, shown in Fig.~\ref{fig:struct}(b), contains 6 f.u. Relation between the conventional and primitive cells is visualized in Supplemental Material~\footnote{See Fig. S1 in Supplemental Material  for the relation between the conventional and primitive cells.}.
Experimental and theoretical~\cite{cabi2} lattice parameters and atomic positions are shown in Table~\ref{tab:cryst}. 
Conventional unit cell is elongated about 3.5 times along the b-axis, comparing to other dimensions. 
This is related to the quasi-two dimensional character of CaBi$_2$ crystal structure, with a sequence of atomic Bi(2) and Bi(1)-Ca layers, perpendicular to the b-axis, which form [Ca-Bi(1)]-[Bi(2)]-[Ca-Bi(1)] ,,sandwiches''. 
This quasi-2D geometry of the system, reflected also in the charge density distribution, was discussed in more details in Ref.~\cite{cabi2}.

Calculations in this work were done using the {\sc Quantum ESPRESSO} software~\cite{QE-2009,QE-2017}, which is based on   
density-functional theory (DFT) and pseudopotential method.
We used RRKJ (Rappe-Rabe-Kaxiras-Joannopoulos) ultrasoft pseudopotentials~\cite{pseudo},
with the PBE-GGA~\cite{pbe} (Perdew-Burke-Ernzerhof generalized gradient approximation) for the exchange-correlation 
potential. 
For bismuth atom, both fully-relativistic and scalar-relativistic pseudopotentials were used, whereas 
for calcium only the scalar-relativistic pseudopotential was taken, as inclusion of SOC in its 
pseudopotential didn't affect the electronic structure of CaBi$_2$. 
At first, unit cell dimensions and atomic positions were relaxed with BFGS (\textit{Broyden-Fletcher-Goldfarb-Shanno}) 
algoritm, where the experimentally determined crystal structure parameters were 
taken as initial values (see, Table~\ref{tab:cryst}). For the relativistic case (with SOC included), the unit cell dimensions were taken from the scalar-relativistic calculations, whereas the atomic positions were additionally relaxed.
Next, the electronic structure was calculated on the Monkhorst-Pack 
grid of $12^3$ {\bf k}-points.
In the following step, the dynamical matrices were computed 
on the grid of $4^3$ {\bf q}-points, using DFPT~\cite{dfpt} (density-functional perturbation theory). 
Through double Fourier interpolation, real-space interatomic-force constants were obtained and used to compute the phonon dispersion relations.  
Finally, the Eliashberg electron-phonon interaction function $\alpha^2F(\omega)$ was calculated 
using  the self-consistent first-order variation of the crystal potential from preceding phonon calculations, where 
summations over the Fermi surface was done using a dense grid of $24^3$ {\bf k}-points. 
Obtained $\alpha^2F(\omega)$ was used to calculate the electron-phonon coupling constant $\lambda$ in both scalar, and relativistic cases, and by using 
the Allen-Dynes equation~\cite{allen-dynes}, critical temperature was determined.

\begin{figure*}[t]
\includegraphics[width=\textwidth]{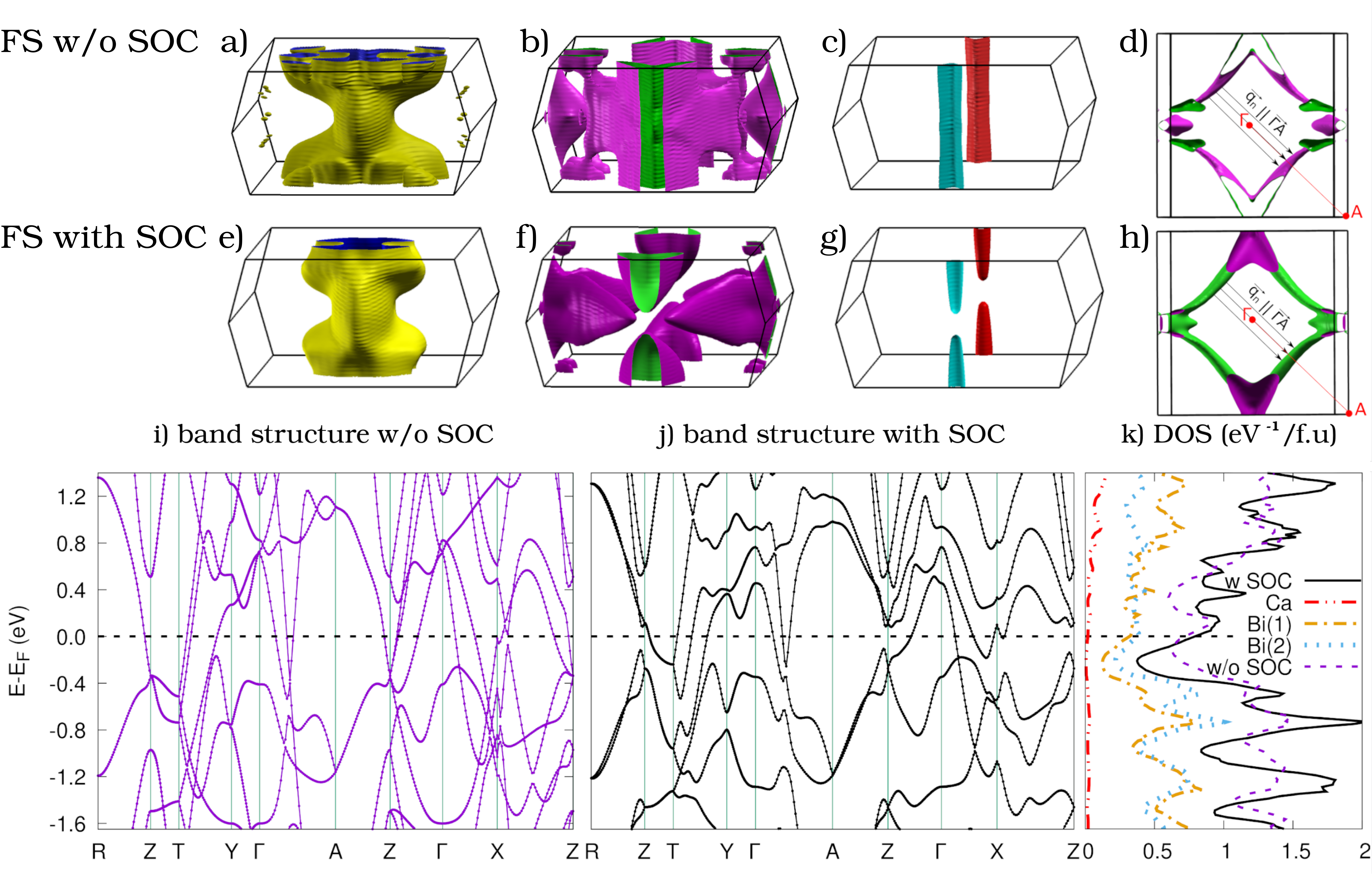}
\caption{The electronic structure of CaBi$_2$ without and with SOC: (a-c) and (e-g) three Fermi surface (FS) pieces; (d), (h) second FS piece with the nesting vector indicated; (i-j) electronic dispersion relations near the Fermi energy ($E_F$); (k) density of states (DOS), with the partial atomic densities plotted only for the relativistic case.}
\label{fig:electron}
\end{figure*}

\begin{table}[t]
\caption{Theoretical and experimental~\cite{cabi2} crystal structure parameters of CaBi$_2$, space group {\it Cmcm}, no. 63. Theoretical values were obtained in scalar-relativistic calculations (w/o SOC) and relativistic calculations (w SOC), where for the latter case only atomic positions were relaxed, with a, b, c taken from the scalar-relativistic relaxation.
All atoms occupy (4c) positions, $(0,y,0.25)$, where $y$ is given below. Conventional unit cell 
parameters are expressed in \AA~units, primitive cell angle $\alpha$ is shown in Fig~\ref{fig:struct}.}
\label{tab:cryst}
\begin{center}
\begin{ruledtabular}
\begin{tabular}{ l c c c c c  c c}
  & 	a& b & c & $\alpha$ & $y$-Ca & $y$-Bi(1) & $y$-Bi(2) \\
\hline
expt.	        & 4.696	&17.081	& 4.611 & $30.74^\circ$ &  0.4332& 0.0999 & 0.7552\\
w/o SOC	& 4.782	&17.169	& 4.606 & $31.16^\circ$ &  0.4015& 0.0655 &0.7575\\
w SOC	&     	&   	&       &           &  0.4006& 0.0668 &0.7555\\

\end{tabular}
\end{ruledtabular}
\end{center}
\end{table}

\section{Electronic structure}

Electronic structure of CaBi$_2$ has been initially presented in Ref.~\cite{cabi2}, 
however for the sake of clarity and consistency of the present work it is briefly discussed also here.
Fig.~\ref{fig:electron} shows electronic dispersion relations, densities of states (DOS) and Fermi surface (FS) of CaBi$_2$.
Brillouin zone of the system, with location of high-symmetry points, is shown in Fig.~\ref{fig:struct}(d). 

Fig.~\ref{fig:electron} shows both scalar- and full-relativistic results, to visualize the influence of SOC on the electronic structure.
As already mentioned~\cite{cabi2}, the studied system has a layered structure, 
with metallic Bi(2) layers and more ionic Ca-Bi(1) layers, stacked in [Ca-Bi(1)]-[Bi(2)]-[Ca-Bi(1)] ,,sandwiches'' along the $b$-axis.
This is reflected in the computed band structure, which is generally less-dispersive for the $k_y$ direction, parallel to the $b$-axis 
(see bands e.g. in $\Gamma$-Y and Z-T directions), 
and more dispersive in others.

Three bands are crossing the Fermi level and form three pieces of Fermi surface, plotted in Fig.~\ref{fig:electron}(a)-(d) for 
the scalar-relativistic case, and in Fig.~\ref{fig:electron}(e)-(h) for the relativistic case. 
In general, the quasi-two dimensional structure of the system is seen in the topology of its Fermi surface, 
with the highlighted $k_y$ direction, parallel to the real-space $b$ axis, and perpendicular to atomic layers. 
In line with this, first piece [panels (a) and (e)] is cylindrical along $k_y$.  
The second piece [panels (b) and (f)] is large and rather complex, but also with a reduced dimensionality -- there are large and flat FS areas parallel to $k_y$, while calculated without SOC. 
As there is a special ${\bf q}_n$ vector, which connects flat areas of this part of Fermi surface, as shown in Fig.~\ref{fig:electron}(d), this FS sheet  exhibits strong nesting.
The shortest nesting vector ${\bf q}_n$, which lies in the $\Gamma$-A direction, is about ${3\over 4}$ of $\Gamma$A long, 
however, as seen in Fig.~\ref{fig:electron}(d), nesting condition is also fulfilled for vectors longer than {\bf q}$_n$ shown in the figure.
Also, similar nesting condition is fulfilled for the $\Gamma$-A$_1$ direction, perpendicular to $\Gamma$-A.
{This piece of Fermi surface is most strongly influenced by the spin-orbit interaction, which splits it into separate sheets, considerably reducing the area of its flat parts. 
Thus, SOC reduces the quasi-two-dimensional character of FS and nesting becomes much weaker.}
Changes in topology of this FS piece are caused by the significant shift of the band along T-Z 
and opening of a gap around Z-point, seen in Fig.~\ref{fig:electron}(i)-(j).
Presence of the spin-orbital dependent Fermi surface nesting {will have} strong implications for the electron-phonon interaction, as will be discussed below.
The third, smallest piece of Fermi surface, plotted in Fig.~\ref{fig:electron} (c) and (g), is also strongly two-dimensional and is changed by SOC in a similar way as the second one -- without SOC it is nearly cylindrical along TZ direction (with no dispersion in $k_y$), while calculated with SOC, due to gap opening at Z-point, it is split into two cones.

\begin{figure}[t!]
\includegraphics[width=0.45\textwidth]{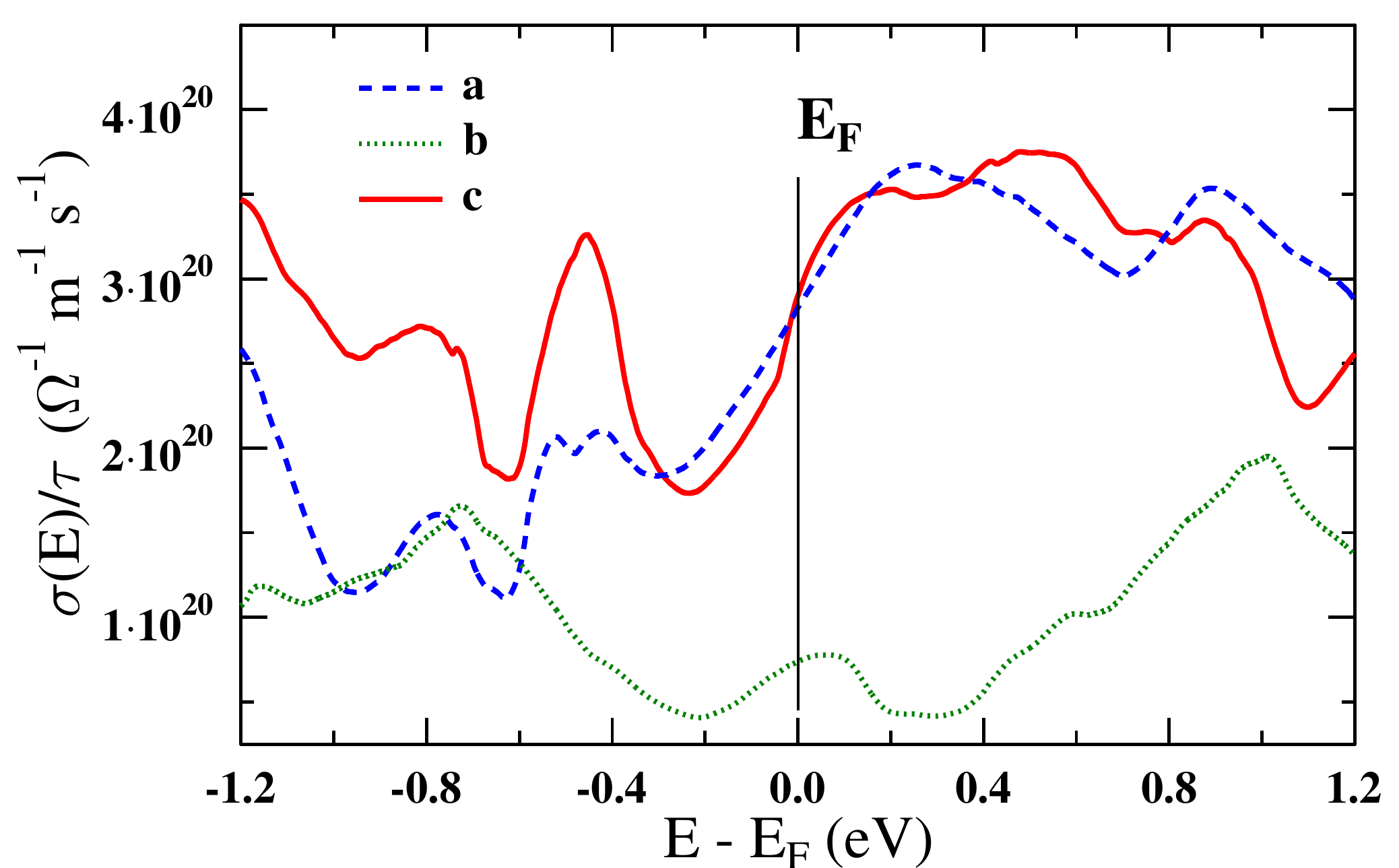}
\caption{\label{fig:trans} Transport function of CaBi$_2$ computed in the constant scattering time approximation, along three unit cell directions.}
\end{figure}

The DOS plot in Fig.~\ref{fig:electron}(k) clearly shows the main role of bismuth atoms in determining electronic properties of CaBi$_2$, 
as most electronic states around the Fermi level originate from bismuth 6p orbitals.
SOC visibly modifies DOS as well, however, as far as the $N(E_F)$ value is concerned, the difference is not substantial, since $N(E_F) = 1.15$ eV$^{-1}$ (with SOC) and $N(E_F) = 1.10$ eV$^{-1}$ (without SOC).

To quantitatively investigate the quasi-2D electronic properties of CaBi$_2$, the electronic transport function of CaBi$_2$ was additionally computed,
within the Boltzmann approach in the constant scattering time approximation (CSTA) and using the {\sc BoltzTraP} code~\cite{boltztrap} .
Fig.~\ref{fig:trans} shows the diagonal elements of the energy dependent electrical conductivity tensor of CaBi$_2$ (transport function $\sigma(E)$).
For each band $i$ and wave vector ${\bf k}$ electrical conductivity is determined by the carrier velocity $v$ and scattering time $\tau$ via $\sigma_{\alpha\beta}(i,{\bf k}) = e^2\tau v_{\alpha}(i,{\bf k})v_{\beta}(i,{\bf k})$. Electron velocities are related to the gradient of dispersion relations $E_i({\bf k})$, $v_{\alpha}(i,{\bf k}) = \hbar^{-1}\partial E_i({\bf k})/\partial k_{\alpha}$, 
thus in the CSTA, by taking $\tau = {\rm const.}$, one may compute $\sigma(E)/\tau$. The diagonal elements of $\sigma_{\alpha\beta}(i,{\bf k})$, integrated over the isoenergy surfaces, are shown in Fig.~\ref{fig:trans} as $\sigma_{\alpha\alpha}(E)/\tau$, where $\alpha = \{a, b, c\}$ are 
the three unit cell directions. 
As one can see, the generally less dispersive band structure along $k_y$ direction in the Brillouin zone is responsible for the smaller electron velocities, making $\sigma(E)$ around $E_F$ about four times smaller along $b$ axis, than in the in-plane $(a,c)$ directions. 

\section{Phonons}

\begin{figure*}[ht!]
\includegraphics[width=\textwidth]{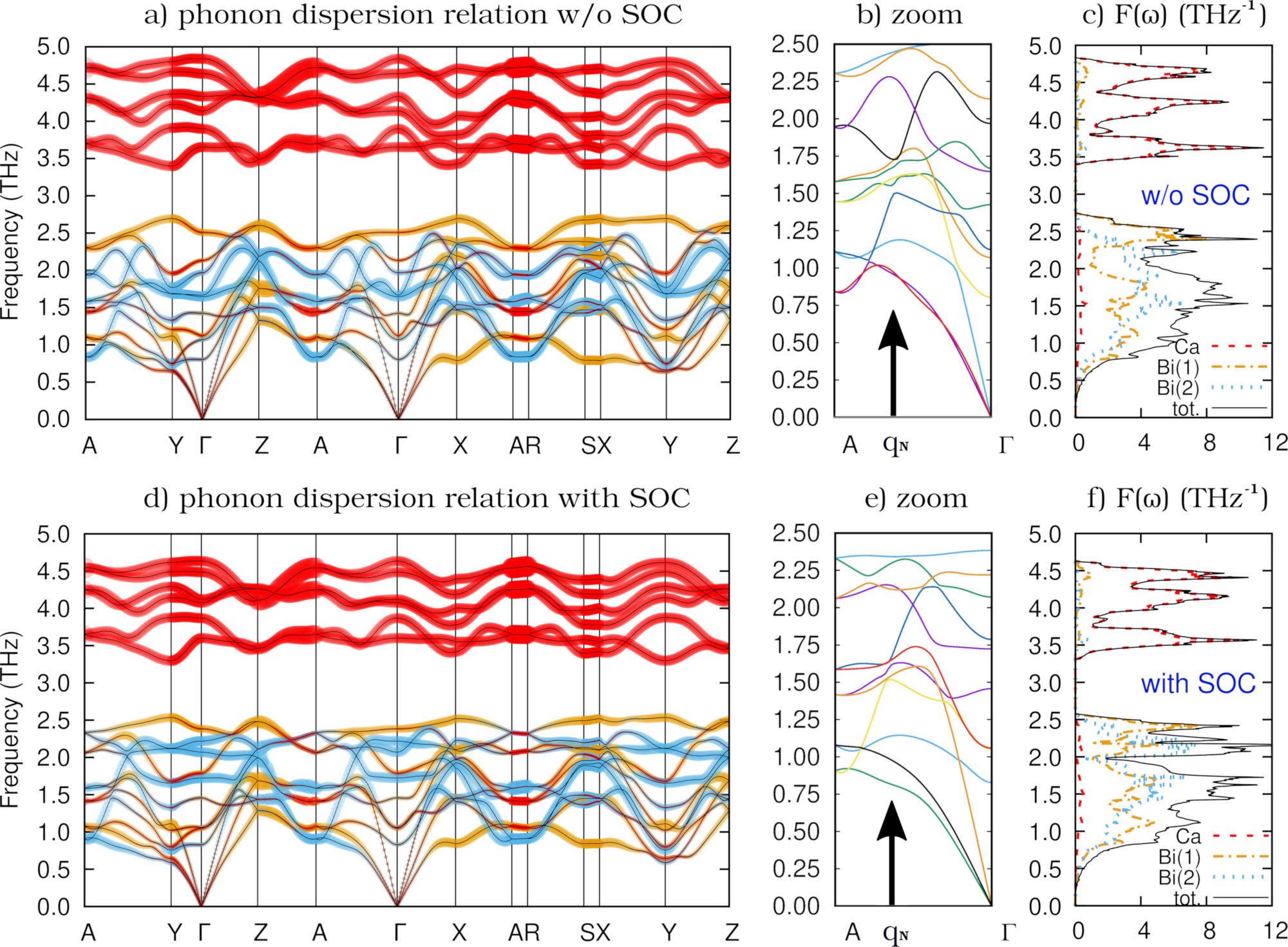}
\caption{Phonon dispersion relations and  phonon DOS $F(\omega)$ of CaBi$_2$ without (at the top) and with (at 
the bottom) SOC. The contributions of atoms in dispersion relations in panels (a) and (d) are marked by colored fat bands; Ca - red, Bi(1) - yellow, Bi(2) - blue. The nesting vector {\bf q}$_n$, shown in Fig.~\ref{fig:electron}, is marked with an arrow in panels (b,e).}
\label{fig:phdos}
\end{figure*}

Figure~\ref{fig:phdos} shows phonon dispersion relations $\omega({\bf q})$ and phonon density of states $F(\omega)$, 
computed without and with SOC. Obtained phonon spectra are stable, i.e. with no imaginary frequencies in both cases.  
As in the primitive cell of CaBi$_2$ there are 6 atoms (2 f.u.), the total number of phonon branches is 18.
Contributions of each of the atom to the phonon branches are marked using colored ''fat bands'', 
additionally partial phonon densities of states are computed.
Due to the large difference in atomic masses ($M_{\rm Bi} \simeq 209$~u, $M_{\rm Ca} \simeq 40$~u) the phonon spectrum is separated into two regions, 
with the low-frequency  
part, dominated by bismuth atoms' vibrations, and high-frequency part, dominated by calcium.
Average total and partial phonon frequencies were computed using the formulas (\ref{eq:mom})--(\ref{eq:omlog2}) given below, and are collected in Table~\ref{tab:freq}.

\begin{equation}\label{eq:mom}
\langle \omega^n \rangle = \int_0^{\omega_{\mathsf{max}}} \omega^{n-1} F(\omega) d\omega \left/ \int_0^{\omega_{\mathsf{max}}} F(\omega) \frac{d\omega}{{\omega}} \right.,
\end{equation}
\begin{equation}\label{eq:sred}
\langle \omega \rangle = \int_0^{\omega_{\mathsf{max}}} \omega F(\omega) d\omega \left/ \int_0^{\omega_{\mathsf{max}}} F(\omega) d\omega \right.,
\end{equation}
\begin{equation}\label{eq:omlog}
\langle\omega_{\rm log}\rangle = \exp\left(\int_0^{\omega_{\mathsf{max}}} F(\omega) \ln\omega\frac{d\omega}{{\omega}} \left/ \int_0^{\omega_{\mathsf{max}}} 
{F(\omega)}\frac{d\omega}{{\omega}} \right. \right),
\end{equation}
%
%
\begin{equation}\label{eq:omlog2}
\langle\omega_{\rm log}^{\alpha^2F}\rangle = \exp\left(\int_0^{\omega_{\mathsf{max}}} \alpha^2F(\omega) \ln\omega\frac{d\omega}{{\omega}} \left/ \int_0^{\omega_{\mathsf{max}}} 
{\alpha^2F(\omega)}\frac{d\omega}{{\omega}} \right. \right).
\end{equation}

\begin{table}[t]
\caption{The phonon frequency moments of CaBi$_2$, computed using Eq.(\ref{eq:mom})--(\ref{eq:omlog2}).}
\label{tab:freq}
\begin{center}
\begin{ruledtabular}
\begin{tabular}{ c c c c c c }
& 
$\langle\omega^1\rangle$ & 
$\sqrt{\langle\omega^2\rangle}$ & 
$\langle\omega\rangle$ & 
$\langle\omega_{\rm log}\rangle$ & 
$\langle\omega_{\rm log}^{\alpha^2F}\rangle$ \\
\hline
\multicolumn{6}{c}{w/o SOC (THz)}\\ 
total         & 1.90 & 2.18 & 2.50 & 1.64 & 1.66 \\
Ca            & 3.52 & 3.71 & 3.91 & 3.20 &  \\
Bi(1)         & 1.55 & 1.70 & 1.87 & 1.40 & \\
Bi(2)         & 1.53 & 1.63 & 1.73 & 1.43 & \\
\multicolumn{6}{c}{with SOC (THz)}\\ 
total         & 1.86 & 2.13 & 2.44 & 1.60 & 1.65 \\
Ca            & 3.47 & 3.65 & 3.83 & 3.16 &  \\
Bi(1)         & 1.49 & 1.62 & 1.77 & 1.36 & \\
Bi(2)         & 1.51 & 1.62 & 1.72 & 1.41 & \\
\end{tabular}
\end{ruledtabular}
\end{center}
\end{table}

Spin-orbit coupling has a visible impact on dynamical properties of CaBi$_2$. At first, SOC 
leads to slightly lower frequencies of phonons, since some of the calcium and bismuth modes are 
shifted towards lower $\omega$. This is seen in phonon frequency moments, collected in Table~\ref{tab:freq}.
However, the gap between the high- and low-frequency group of modes is 
increased, from about 0.7 THz without SOC, to 0.9 THz with SOC, i.e. frequencies of higher Bi modes are influenced 
to a larger degree, than the lower Ca branches. 

\begin{figure}[b]
\includegraphics[width=0.45\textwidth]{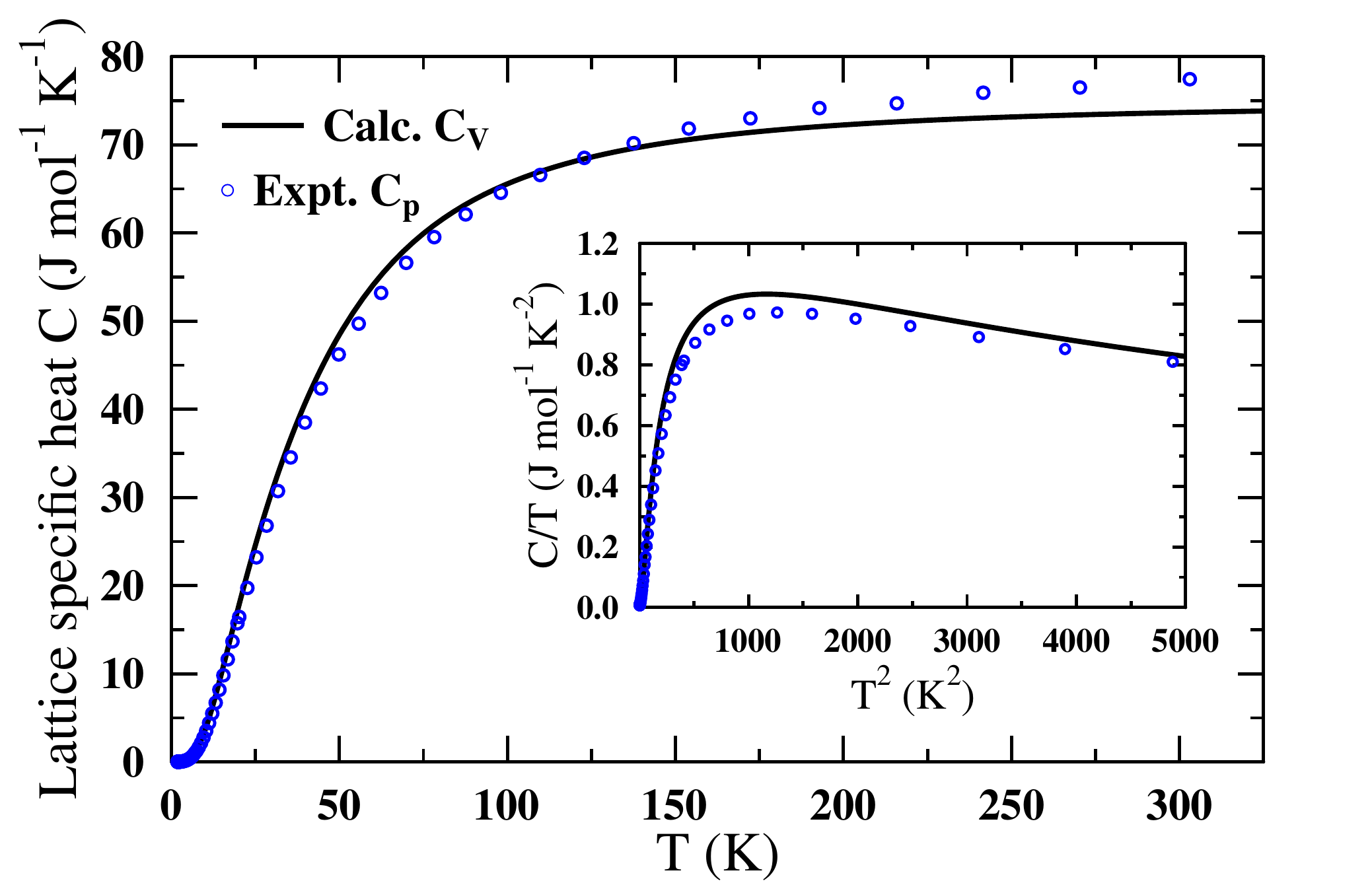}
\caption{Computed constant volume $C_V$ (lines) and measured constant pressure $C_p$~\cite{cabi2} (points) lattice heat capacity of CaBi$_2$, given per formula unit.}
\label{fig:heat}
\end{figure}

Average phonon frequency $\langle \omega \rangle = 2.44$ THz (with SOC) corresponds to temperature of 117~K, lower than the experimentally determined Debye temperature $\Theta_D = 157$~K~\cite{cabi2}. 
As there is no universal definition of the ''theoretical'' Debye temperature for a system with optical phonon branches, to be able to confront our calculations with the experimental findings, constant-volume lattice heat capacity $C_V$ was calculated~\cite{grimvall}:
\begin{equation}
C_V = R\int_0^{\infty}F(\omega)\left(\frac{\hbar\omega}{k_BT}\right)^2\frac{\exp(\frac{\hbar\omega}{k_BT})}{(\exp(\frac{\hbar\omega}{k_BT})-1)^2}
\end{equation}
using the relativistic phonon DOS $F(\omega)$ function. 
In Fig.~\ref{fig:heat} theoretical $C_V$ is compared to the experimental constant-pressure $C_p$ from Ref.~\cite{cabi2} (electronic heat capacity was subtracted from $C_p$) and a good agreement is found.
Deviation at higher temperatures most likely is due the difference of $C_p$ and $C_V$, related to the anharmonic effects, where $C_p \simeq C_V(1+\beta\gamma_GT)$~\cite{grimvall}, where $\beta$ is the volume thermal expansion coefficient and $\gamma_G$ is the Gr\"uneisen parameter.
From the ratio of $C_p/C_V$ at 300 K we can estimate $\beta\gamma_G \simeq 1.7\cdot 10^{-4}$~K$^{-1}$.
At low temperatures, where the difference between $C_p$ and $C_V$ should be small, we observe slightly larger calculated $C_V$, seen better in the $C/T$ {\it vs.} $T^2$ plot in the inset in Fig.~\ref{fig:heat}. Largest difference appears around $T \simeq $30~K, and indicates slightly larger theoretical $F(\omega)$ in the 1 - 2 THz frequency range, than in the real system. However, still the largest differences between experimental and calculated values are of the order of 3-4\%.

In the phonon spectrum, especially in the non-SOC case, we observe Kohn anomalies along $\Gamma$-A direction in Fig.~\ref{fig:phdos}, where some of the phonon frequencies are strongly renormalized and lowered. This part of the spectrum is enlarged in Figs.~\ref{fig:phdos}(b,e) and one observes dips in the phonon branches, as well as the inflection of the acoustic mode, associated with Bi(2) vibrations, in the non-SOC spectrum.
Similar inflection what was observed eg. in palladium~\cite{palladium}. 
In general, the Kohn anomaly~\cite{kohn-anomaly} is an anomaly in the phonon dispersion curve in a metal, 
where the frequency of the phonon is lowered due to screening effects. Such an anomaly appears at the wave vector ${\bf q}_n$
which satisfy the nesting conditions -- when there are flat and parallel parts of Fermi surface, which can be connected by ${\bf q}_n$, 
there are many electronic states which may interact with phonons having the wavevector ${\bf q}_n$.
In CaBi$_2$, as we mentioned above, large parts of Fermi surface sheets, plotted in Figs.~\ref{fig:electron}(b,f), may be connected by the same nesting vector ${\bf q}_n$, which is parallel to $\Gamma$-A direction, as is shown in Fig.~\ref{fig:electron}(d,h) for scalar- and full-relativistic cases, respectively. 
This nesting vector is also marked with an arrow in the dispersion plots in Fig.~\ref{fig:phdos}(b,e).
Since in the scalar-relativistic case much larger parts of this Fermi surface sheet are parallel, nesting is much stronger and thus the anomalies are very pronounced in the non-SOC calculations, as seen in the dispersion plots in Fig.~\ref{fig:phdos}.
The anomaly, observed here near the A-point, will have strong impact on electron-phonon interaction, as will be discussed in the next section.

\section{Electron-phonon coupling }

Electron-phonon interaction can be described in terms of the hamiltonian \cite{grimvall,wierzbowska}: 
\begin{equation}
\hat{H}_{\rm e-p}=\sum_{{\bf k,q},\nu} g_{{\bf q}\nu}({\bf k},i,j) c_{\bf k+q}^{\dagger i}c_{\bf k}^j(b_{{\bf -q}\nu}^\dagger+b_{{\bf q}\nu}),
\end{equation}
The creation and annihilation operators $c_{\bf k+q}^{\dagger i}$, $c_{\bf k}^j$ refer to electrons in the state $\bf k+q$ and $\bf k$ in the $i$-th and $j$-th band respectively, while $b^\dagger_{{\bf -q}\nu}$, $b_{{\bf q}\nu}$ operators describe emission or absorption of the phonon from the $\nu$-th mode and with wave vectors $\bf -q$ or $\bf q$, respectively. The electron-phonon interaction matrix elements $g_{{\bf q}\nu}({\bf k},i,j)$ have the form
\begin{equation}
g_{{\bf q}\nu}({\bf k},i,j) =\left({\hbar\over 2M\omega_{{\bf q}\nu}}\right)^{1/2}
\langle\psi_{i,{\bf k}}| {dV_{\rm SCF}\over d {\hat u}_{\nu} }\cdot
                   \hat \epsilon_{\nu}|\psi_{j,{\bf k}+{\bf q}}\rangle.
\end{equation}
Here $\omega_{{\bf q}\nu}$ is the frequency of $\nu$-th phonon mode at {\bf q}-point, $\psi_{i,{\bf k}}$ is an electron wave function at $\bf k$-point, $\hat{\epsilon}_\nu$ is a phonon polarization vector and ${dV_{\rm SCF}\over d {\hat u}_{\nu}}$ is a change of electronic potential, calculated in the self-consistent cycle, due to the displacement of an atom, $\hat{u}_\nu$. 
On this basis one can calculate the phonon linewidth
\begin{equation}
\begin{split}
\gamma_{{\bf q}\nu} =& 2\pi\omega_{{\bf q}\nu} \sum_{ij}
                \int {d^3k\over \Omega_{\rm BZ}}  |g_{{\bf q}\nu}({\bf k},i,j)|^2 \\
                    &\times\delta(E_{{\bf q},i} - E_F)  \delta(E_{{\bf k+q},j} - E_F), 
\end{split}
\end{equation}
where $E_{{\bf k},i}$ refers to the energy of an electron.
The phonon linewidth describes the strength of the interaction of the electron at the Fermi surface with the phonon from the $\nu$-th mode, which has the wave vector $\bf q$, and it is inversely proportional to the lifetime of the phonon. 
Now, the Eliashberg function can be defined as
\begin{equation}
\alpha^2F(\omega) = {1\over 2\pi N(E_F)}\sum_{{\bf q}\nu} 
                    \delta(\omega-\omega_{{\bf q}\nu})
                    {\gamma_{{\bf q}\nu}\over\hbar\omega_{{\bf q}\nu}}.
\end{equation}
where $N(E_F)$ refers to the electronic DOS at Fermi level.
The Eliashberg function is proportional to the sum over all phonon modes and all $\bf q$-vectors of phonon linewidths divided by their energies, and  describes the interaction of electrons from the Fermi surface with phonons having frequency $\omega$.
The total electron-phonon coupling parameter $\lambda$ may be now defined using the $\alpha^2F(\omega)$ function, as:
\begin{equation}\label{eq:lam2}
\lambda=2\int_0^{\omega_{\rm max}} \frac{\alpha^2F(\omega)}{\omega} \text{d}\omega,
\end{equation}
or alternatively, directly by the phonon linewidths: 
\begin{equation}\label{eq:lam}
\lambda = \sum_{{\bf q},\nu}\frac{\gamma_{{\bf q}\nu}}{\pi \hbar N(E_F)\omega^2_{{\bf q},\nu}}.
\end{equation}
More detailed description of the theoretical aspects of the electron-phonon coupling can be find in~\cite{grimvall,wierzbowska}.

\begin{table}[t!]
\caption{Frequencies $\omega_{\nu{\bf q}}$ (THz) and linewidths $\gamma_{\nu{\bf q}}$ (GHz) of eighteen, doubly-degenerated phonon modes $\nu$ at the ${\bf q}$-point A, obtained in scalar-relativistic (scalar) and relativistic (rel) calculations.}
\label{tab:freqA}
\begin{center}
\begin{ruledtabular}
\begin{tabular}{ l c c c c c c c c c}
$\nu$  & 1-2 & 3-4 & 5-6 & 7-8 & 9-10 & 11-12 & 13-14 & 15-16 & 17-18 \\
\hline
$\omega_{\rm scalar}$   &0.84 & 1.10 & 1.44 & 1.57 & 1.93 & 2.29 & 3.68 & 4.29 & 4.69\\
$\omega_{\rm rel}$  &0.91 & 1.08 & 1.42 & 1.59 & 2.06 & 2.33 & 3.65 & 4.25 & 4.54\\
$\gamma_{\rm scalar}$   & 1.9 & 39.4 & 67.5 & 13.3 & 8.3 & 48.9  & 29.7 & 37.9 & 15.8\\
$\gamma_{\rm rel}$      &0.6  & 0.5 & 0.3   & 0.8 & 2.0  & 1.9   & 0.9  & 0.9  & 1.2 
\end{tabular}
\end{ruledtabular}
\end{center}
\end{table}

\begin{figure*}[htb!]
\begin{flushleft}
\includegraphics[width=1\textwidth]{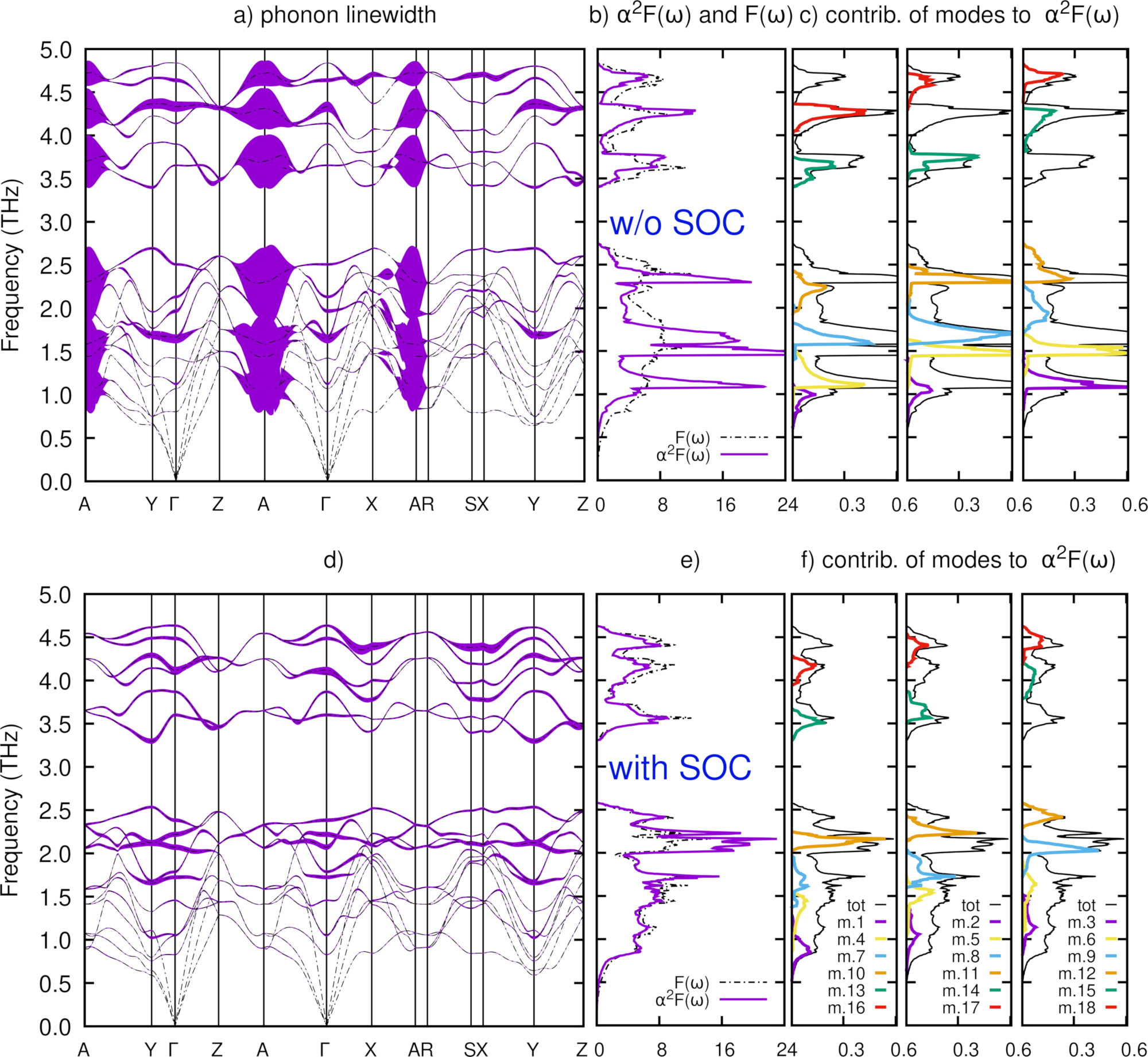}
\end{flushleft}
\caption{Electron-phonon coupling in CaBi$_2$; (a-c) without SOC, (d-f) with SOC. 
Phonon dispersion relations with shading corresponding to phonon linewidth $\gamma_{{\bf q}\nu}$ [panels (a,d)]. In both panels (a,d), $\gamma_{{\bf q}\nu}$ (in THz units) is multiplied by 4, to make it visible for the SOC case. Panels (b,e) show the Eliashberg function $\alpha^2F(\omega)$, renormalized to $3n$ ($n=6$ is the number of atoms in the primitive cell), with the phonon DOS $F(\omega)$ plotted in the background; panels (c,f) show the actual $\alpha^2F(\omega)$ with decomposition over all the 18 phonon modes.}
\label{fig:a2f}
\end{figure*}

Figures~\ref{fig:a2f}(a,d) display the phonon dispersion curves, with shading corresponding to the phonon linewidth $\gamma_{{\bf q}\nu}$ (in THz units) for the mode $\nu$ at ${\bf q}$-point. To make $\gamma_{{\bf q}\nu}$ visible for the SOC case, $\gamma_{{\bf q}\nu}$ is multiplied by 4, and the same multiplicator is kept on both panels (a,d) to ensure the same visual scale.
Eliashberg function $\alpha^2F(\omega)$, plotted on the top of phonon DOS, $F(\omega)$, is shown in panels (b,e), and $\alpha^2F(\omega)$ decomposed over the 18 phonon modes in panels (c,f). 
In panels (b,e), Eliashberg function is renormalized to $3n$ ($n$ - number of atoms in the primitive cell), in the same way as phonon DOS, to allow for a direct comparison of both functions.
Each of the quantities is plotted as obtained from scalar-relativistic calculations [Fig.~\ref{fig:a2f}(a,b,c)], and full-relativistic calculations [Fig.~\ref{fig:a2f}(e,f,g,h)].
The finite width of the phonon lines, according to Eq.(\ref{eq:lam}), is a measure of a local strength of the electron-phonon interaction.
One thing that immediately catches the eye is a huge phonon linewidth $\gamma_{{\bf q}\nu}$ around the A-point in the scalar-relativistic results in Fig.~\ref{fig:a2f}(a). 
This large $\gamma_{{\bf q}\nu}$ area starts at the nesting vector ${\bf q}_n$ and is related to the presence of the Kohn anomaly and Fermi surface nesting. The large number of electronic states, which may interact with phonons having wavevectors from this area of the Brillouin zone, makes the electron-phonon interaction strong and anisotropic.
Comparing Fig.~\ref{fig:a2f}(a) and Fig.~\ref{fig:phdos}(a) we also see, that the strong electron-phonon interaction around the A point is related to the Ca and Bi(1) atoms vibrations, with much smaller contribution from Bi(2) atomic modes. These strong-coupling modes involve both in-plane and out-of-plane Ca and Bi(1) atomic displacements, as can be seen in the displacement patterns shown in Supplemental Material~\footnote{See Fig. S2 in Supplemental Material  for the phonon displacement patterns in A point.}, however the corresponding phonon wave vectors are confined to the in-plane $q_x-q_z$ directions. This is correlated with the quasi-2D layered structure of this compound and shows signatures of the two-dimensional character of the electron-phonon interaction here.
Frequencies and phonon linewidths $\gamma_{{\bf q}\nu}$ of all doubly-degenerated phonon modes in A-point are shown in Table~\ref{tab:freqA}.

\begin{table*}[htb!]
\caption{Contributions to the total electron-phonon coupling constant $\lambda$ from each of the eighteen phonon branches of CaBi$_2$, the total $\lambda = \sum_{\nu} \lambda_{\nu}$.}
\label{tab:lambda-modes}
\begin{ruledtabular}
\begin{tabular}{ l c c c c c c c c c c c c c c c c c c c c }
mode no. & 	1&	2&	3&	4&	5&	6&	7&	8&	9&	10&	11&	12&	13&	14&	15&	16&	17&	18 &total $\lambda$\\
\hline
$\lambda_{\nu}$ w/o SOC &	0.04&	0.05&	0.10&	0.10&	0.09&	0.09&	0.07&	0.11&	0.04&	0.03&	0.05&	0.04&	0.02&	0.02&	0.02&	0.03&	0.01&	0.01 & 0.94\\
$\lambda_{\nu}$ w SOC &	0.06&	0.04&	0.04&	0.04&	0.04&	0.04&	0.03&	0.04&	0.05&	0.04&	0.03&	0.02&	0.01&	0.01&	0.01&	0.01&	0.01&	0.01 & 0.54
\end{tabular}
\end{ruledtabular}
\end{table*}

\begin{table}[b]
\caption{Electron-phonon coupling constant $\lambda$ and critical temperature $T_c$ calculated for 
CaBi$_2$ without SOC [w/o SOC]; with SOC [w SOC]; extracted from the experimental data using $T_c$ and the McMillan formula [expt. ($T_c$)]; extracted from the experimental data using electronic heat capacity coefficient $\gamma$ and theoretical $N(E_F)$ with SOC [ expt. ($\gamma$)].}
\label{tab:lambda}
\begin{center}
\begin{ruledtabular}
\begin{tabular}{ c c c c c }
& w/o SOC & w SOC &  expt. ($T_c$) & expt. ($\gamma$) \\
\hline
$N(E_F)$ (eV$^{-1}$) & 1.10 & 1.15 &   & \\
$\lambda$& 0.94 & 0.54 &  0.53 & 0.51\\
$T_c$ (K)& 5.2 & 1.3 & 2.0 &  
\end{tabular}
\end{ruledtabular}
\end{center}
\end{table}

The strong anisotropy and mode-dependence of the electron-phonon interaction in CaBi$_2$ in the scalar-relativistic case results in the Eliashberg function having significantly different shape, than the phonon DOS function $F(\omega)$, as seen in Fig.~\ref{fig:a2f}(b). $\alpha^2F(\omega)$ is strongly peaked around the seven frequencies of phonon modes from the A-point, which have large $\gamma_{{\bf q}\nu}$. Contributions of each of the total number of 18 phonon modes to the total $\alpha^2F(\omega)$ function are plotted in Fig.~\ref{fig:a2f}(c), and $\lambda_{\nu}$ values are collected in Table~\ref{tab:lambda-modes}.
Total electron-phonon coupling constant is directly calculated from the Eliashberg function, using Eq.(\ref{eq:lam2}),
which gives $\lambda = 0.94$. This value is considerably larger, than expected from the experimental value of $T_c$ via inverted McMillan formula~\cite{mcmillan} $\lambda = 0.53$. The latter value is calculated using the experimental Debye temperature $\Theta_D = 157$~K~\cite{cabi2}, and assuming Coulomb pseudopotential parameter $\mu^* = 0.10$, since CaBi$_2$ is a simple metal with $s$ and $p$ electrons and low $N(E_F)$ value~\footnote{It is worth recalling here, that McMillan~\cite{mcmillan} for his $T_c$ formula assumed $\mu^* = 0.13$ for transition metals and $\mu^* = 0.10$ for simple metals, whereas Allen and Dynes~\cite{allen-dynes} recommended using their formula with $\mu^* = 0.10$ for transition metals, and even lower values, like $\mu^* = 0.09$, for simple metals. We consistently use $\mu^* = 0.10$ here. Taking $\mu^* = 0.13$  with McMillan formula and experimental $T_c$ for CaBi$_2$ gives slightly larger $\lambda_{\rm expt} = 0.59$.}.
Also, $\lambda$ may be extracted in an usual way from the experimental value of the electronic heat capacity coefficient 
$\gamma_{\rm expt}$= 4.1~mJ/(mol K$^2$) and calculated $\gamma_{\rm calc} = \frac{\pi^2}{3}k_B^2N(E_F)$, where $k_B$ is the Boltzmann constant and $N(E_F)$ is the DOS at the Fermi level, if one assumes that the measured Sommerfeld coefficient $\gamma_{\rm expt}$ is renormalized by the electron-phonon interaction only:
\begin{equation}\label{eq:gamma}
\lambda=\frac{\gamma_{\rm expt}}{\gamma_{\rm calc}} -1.
\end{equation}
This gives $\gamma_{\rm calc} = 2.59$~mJ/(mol K$^2$) and similar value of $\lambda = 0.58$, much smaller, than $\lambda = 0.94$ obtained in the scalar-relativistic calculations.

When spin-orbit coupling is included, however, due to the change in Fermi surface shape and reduction of the area of flat parts of FS, connected by the nesting vector ${\bf q}_n$ [see, Fig.~\ref{fig:electron}(b,f)], the overall strength of the electron-phonon interaction is reduced, both in relation to the A-point area, and to the total $\lambda$. 
As can be seen in Figs.~\ref{fig:a2f}(e,f), in this case electron-phonon interaction becomes less mode- and ${\bf q}$-dependent, and huge $\gamma_{{\bf q}\nu}$ around A-point are absent. From the values of the phonon frequencies and linewidths at the A-point, collected in Table~\ref{tab:freqA}, we precisely see the strong impact of SOC on the electron-phonon interaction: relatively small changes in phonon frequencies $\omega$ are 
followed by a reduction of $\gamma_{{\bf q}\nu}$ by a factor 10 to 100. 
As now the coupling of electrons to those planar phonon modes is not enhanced any more, in the relativistic case the electron-phonon interaction is more three-dimensional and weakly depends on frequency, and thus the Eliashberg function now closely follows the phonon DOS $F(\omega)$ function, as presented in Fig.~\ref{fig:a2f}(e). The relative enhancement of the electron-phonon coupling occurs for the last three optic modes from the lower-frequency part of the spectrum before the gap. These are modes no. 10, 11 and 12 in Fig.~\ref{fig:a2f}(f), 
located between 2.0 and 2.2 THz. Atomic displacement patterns for these modes in Y point, where the phonon linewidths are relatively large, are shown in Supplemental Material~\footnote{See Fig. S3 in Supplemental Material  for the phonon displacement patterns in Y point.}. In the mode no. 12 we find Bi(1) and Ca vibrations perpendicular to atomic layers, whereas in modes 11 and 10 mostly Bi(2) atoms are involved in the in-plane vibrations. 
Due to overlap of these three modes in 2.0-2.2 THz frequency range, coupling is here enhanced and $\alpha^2F(\omega)$ is above the bare DOS function $F(\omega)$, if both are normalized to the same value (3$n$ in Fig.~\ref{fig:a2f}(e)). But if we take a look at Table~\ref{tab:lambda-modes}, due to strong dependence of $\lambda$ also on the phonon frequency in Eq.(\ref{eq:lam}), $\lambda \propto \omega^{-2}$, the largest contributions per phonon mode come from the $1^{\rm st}$, lowest acoustic mode and from the $9^{\rm th}$ optic mode, which involves Ca and Bi(1) vibrations. 

\begin{figure}[b]
\includegraphics[width=0.45\textwidth]{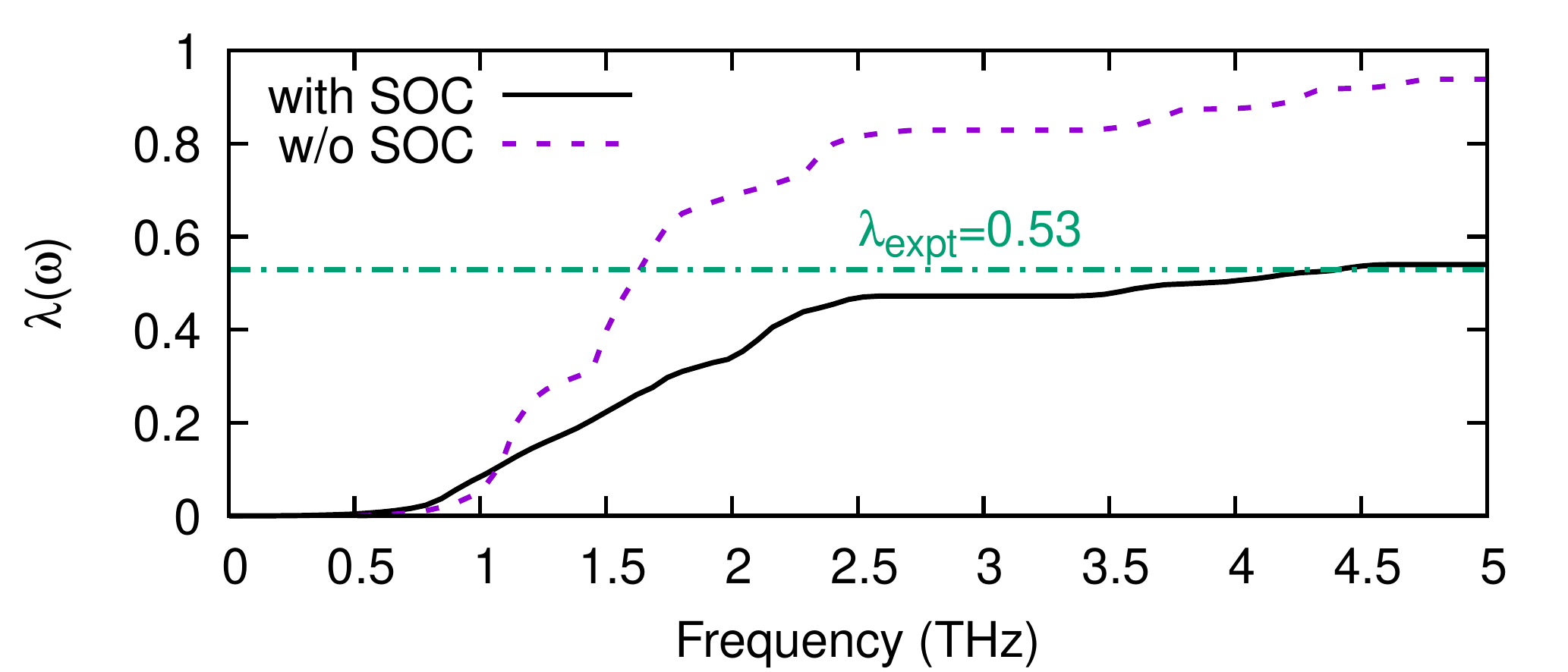}
\caption{The cummulative frequency distribution of $\lambda$, defined as: $\lambda(\omega) = 2\int_0^{\omega} \alpha^2F(\Omega) {d\Omega\over \Omega}$, 
         for scalar and relativistic cases.}
\label{fig:lomega}
\end{figure}

The cummulative frequency distribution of $\lambda$ is shown in Fig.~\ref{fig:lomega}. For both, scalar and relativistic case, main contribution to the electron-phonon coupling constant comes from the phonon modes, located between 1.0 and 2.5 THz. For the scalar-relativistic case, $\lambda(\omega)$ has three steps, due to peaks in the Eliashberg function, appearing before the gap of the phonon spectrum in Fig.~\ref{fig:a2f}. 
When the spin-orbit coupling is included, as has been mentioned above, electron-phonon interaction becomes less frequency-dependent, thus $\lambda(\omega)$ is nearly a linear function in this frequency range. As $\omega$ increases above the gap, the relative contribution of the higher energy modes to $\lambda$ become small, as almost 90\% of the total $\lambda$ is provided by phonons with $\omega < 2.5$~THz.

\begin{table*}[htb!]
\begin{ruledtabular}
\caption{Experimental and theoretical data for selected binary Bi-based superconductors. $\lambda_{\rm calc}$ and $T_c$ are calculated from DFT in a scalar-relativistic (scalar) and relativistic (rel) way; $\lambda(T_c)$ is calculated from experimental $T_c$, McMillan formula, Debye temperature $\Theta_D$ and taking $\mu^* = 0.10$; $\lambda(\gamma_{\rm expt})$ is computed using Eq.(\ref{eq:gamma}) from experimental Sommerfeld parameter $\gamma_{\rm expt}$ and theoretical "bare" value of $N(E_F)$, as given in referenced literature.}
\label{tab:rev}
\begin{tabular}{ l c c c c c c  c c c c c}
 & Space group &$\lambda_{\rm calc}^{\rm scalar}$ & $\lambda_{\rm calc}^{\rm rel}$ & $\gamma_{\rm expt}$ & $\lambda(T_c)$ &$\lambda(\gamma_{\rm expt})$ & $T_{c}^{\rm scalar}$ & $T_{c}^{\rm rel}$ & $T_{c}^{\rm expt}$ & $\Theta_D$ & Ref.\\ 
 & &  &&$(\frac{\text{mJ}}{\text{mol}\text{K}^2}$)&      &       & (K)  & (K)  &  (K)   & (K)   & \\
\hline
CaBi$_2$ &  $CmCm$      & 0.94  & 0.54 & 4.1  & 0.53 & 0.51  & 5.2  & 1.3  & 2.0   & 157  & this work, \cite{cabi2}\\ 
KBi$_2$  & $Fd\bar{3}m$ & 0.76  &      & 1.3  & 0.70 &  0.7  & 2.73 &      & 3.6   & 123  & \cite{kbi2-elph}\cite{kbi2-exp}\\ 
NaBi     & $P4/mmm$     &       &      & 3.4  & 0.56 & 1.05  &      &      & 2.1   & 140  & \cite{nabi}\\ 
BaBi$_3$ & $Fd\bar{3}m$ &       &      & 3.2  & 0.76 & -0.43 &      &      & 6.0   & 171  &  \cite{abi3-b}\\ 
         &              &       & 1.43 & 41   & 0.83 &  6.25 &      & 5.29 & 5.9   & 142  & \cite{abi3,babi3-h}\\ 
         &              &       &      & 49.2 & 0.81 &  7.70 &      &      & 5.9   & 149  &  \cite{abi3-exp-j}\\ 
SrBi$_3$ & $Pm\bar{3}m$ & 0.91  & 1.1  & 14   & 0.91 & 1.74  & 3.73 & 5.15 & 5.5   & 111  & \cite{abi3,srbi3-k}\\ 
         &              &       &      & 6.5  & 0.72 & 0.27  &      &      & 5.6   & 180  &  \cite{abi3-b}\\ 
         &              &       &      & 11   & 0.72 & 1.16  &      &      & 5.5   & 180  & \cite{abi3-exp-j}\\  
LaBi$_3$ & $Pm\bar{3}m$ & 0.90  & 1.35 &      &      &       & 3.71 & 6.88 & 7.3   &      & \cite{labi3,labi3-elph}\\ 
CaBi$_3$ & $Pm\bar{3}m$ & 1.23  &      &      &      &       & 5.16 &      & 1.7   &      & \cite{cabi3,abi3-exp-1st}\\ 
CoBi$_3$ & $Pnma$       &       &      & 16.7 & 0.41 & 1.19  &      &      & 0.5  & 124  & \cite{cobi3}\\ 
NiBi$_3$ & $Pnma$       &       &      & 12.7 & 0.70 & 1.45  &      &      & 4.1   & 141  &\cite{nibi3,nibi3-j}\\ 
         &              &       &      & 11.1 & 0.73 & 1.14  &      &      &       & 128  & \cite{nibi3,nibi3-i}\\ 
\end{tabular}
\end{ruledtabular}
\end{table*}

Now, moving to the total electron-phonon coupling parameter, in the relativistic case $\lambda = 0.54$, which is now in an excellent agreement with the above-mentioned values, determined from the experimental $T_c$ ($\lambda = 0.53$), as well as from the Sommerfeld parameter renormalization factor ($\lambda = 0.51$, when taking the relativistic $N(E_F)$ value). These numbers are summarized in Table~\ref{tab:lambda}.

Using the calculated $\lambda$ and $\langle\omega_{\rm log}^{\alpha^2F}\rangle$ values, and the Allen-Dynes~\cite{allen-dynes} formula:
\begin{equation}\label{eq:tc}
k_{B}T_{c}=\frac{\hbar\langle\omega_{\rm log}^{\alpha^2F}\rangle}{1.20}\,
\exp\left\{-\frac{1.04(1+\lambda)}{\lambda-\mu^{\star}(1+0.62\lambda)}\right\}
\end{equation}
superconducting critical temperatures are calculated and included in Table~\ref{tab:lambda}. 
The Coulomb pseudopotential parameter was kept at $\mu^{\star} = 0.10$.
As in the case of $\lambda$, in the scalar-relativistic calculations obtained value of the critical temperature $T_c = 5.2$~K is 
considerably above the experimental $T_c = 2.0$~K.
Better agreement with experiment is reached after including the spin-orbit coupling, as it lowers computed $T_c$ to 1.3 K, only slightly lower than the experimental one. 

Our calculations show that in CaBi$_2$ the spin-orbit coupling has a very strong and detrimental effect on the electron-phonon interaction and superconductivity. 
{This effect is indirect here, as is caused by the reduction of the Fermi surface nesting, which leads to important changes in $\omega$- and ${\bf q}$-dependence of the electron-phonon interaction. As a result, which SOC electron-phonon interaction is more three-dimensional and isotropic, comparing to the scalar-relativistic case.} SOC effectively weakens the electron-phonon coupling by 42\%: from $\lambda_{\rm scalar} = 0.94$ to $\lambda_{\rm rel} = 0.54$. 
This underlines the need of including SOC in calculations of the electron-phonon coupling in compounds based on such heavy elements, like bismuth, where SOC strongly affects Fermi surface of the material.

In Table~\ref{tab:rev} we have gathered available computational and experimental data on a number of related binary intermetallic superconductors containing Bi. 
Comparing our results to those reported recently for $A$Bi$_3$ ($A$ = Ba, Sr, La) we notice, that SOC effect on the electron-phonon interaction and superconductivity in CaBi$_2$ is stronger, if the relative change between the calculated $\lambda_{\rm scalar}$ and $\lambda_{\rm rel}$ is taken as an indicator. 
Moreover, in CaBi$_2$ the effect is opposite, since in $A$Bi$_3$ SOC enhances the electron-phonon interaction, $\lambda$ and $T_c$.
What is worth noting here, except for CaBi$_2$, there are large differences in $\lambda$ values, obtained from experimental $T_c$ via McMillan equation, and from the Sommerfeld electronic heat capacity coefficient and computed $N(E_F)$ values (Eq.~\ref{eq:gamma}). For the two cases of KBi$_2$ and BaBi$_3$, the computed $\gamma_{\rm calc}$ values are even larger than the measured $\gamma_{\rm expt}$, making $\lambda$ negative, and showing that those systems require reinvestigation. 
Especially BaBi$_3$, where two other reported values of $\gamma_{\rm expt} > 40$~mJ/(mol K$^2$) are large beyond expectations, and also result in spurious values of $\lambda \sim 6-7$~\footnote{Sommerfeld coefficient $\gamma_{\rm expt}$ may be renormalized by other effects than the electron-phonon interaction, however in such intermetallic compounds, with $s$ and $p$ electrons at the Fermi level, strong electron correlactions or paramagnons are not expected to appear.}. 

\section{Summary and conclusions}

First principles calculations of the electronic structure, phonons and the electron-phonon coupling function have been reported for the intermetallic superconductor
CaBi$_2$. 
Calculations were performed within the scalar-relativistic (without the spin-orbit coupling) and relativistic (with spin-orbit coupling) approach, which allowed us to discuss the SOC effect on the computed physical properties.
Electronic structure and electronic transport function reflect the quasi-2D layered structure of the studied compound. 
Dynamic spectrum of CaBi$_2$ is separated into two parts, dominated by the heavier (Bi) and lighter (Ca) atoms' vibrations. 
Strong influence of SOC on the electron-phonon interaction was found. 
In the scalar-relativistic case, due to strong nesting between the flat sheets of the Fermi surface and presence of a large Kohn anomaly, electron-phonon interaction is enhanced in the vicinity of the A point in the Brillouin zone. 
This enhancement of the electron-phonon interaction has a two-dimensional character, as electrons from the flat parts of FS are strongly coupled to phonons, propagating in $q_x-q_z$ directions, and which involve displacement of atoms from the Ca-Bi(1) layers.
{When SOC is included, however, due to the change in Fermi surface topology, nesting becomes weaker and the electron-phonon coupling becomes more isotropic and less $\omega-$dependent. As a result, SOC reduces the magnitude of the electron-phonon coupling in about 42\%, from $\lambda_{\rm scalar} = 0.94$ to $\lambda_{\rm rel} = 0.54$, in an opposite way to related ABi$_3$ superconductors. }
Critical temperature, calculated using the Allen-Dynes equation and relativistic electron-phonon coupling constant gives $T_c = 1.3$~K.
The computed relativistic values of $\lambda$ and $T_c$ remain in a good agreement with experimental results, where $T_c = 2.0$~K and $\lambda = 0.51$ (from the Sommerfeld parameter renormalization) or $\lambda = 0.53$ (from $T_c$, $\Theta_D$ and McMillan equation).
Our results confirm the need of including the spin-orbit coupling in calculations of the electron-phonon interaction functions for materials containing such heavy elements, like Bi, and where SOC strongly modifies Fermi surface of the system.
Finally we may summarize, that CaBi$_2$ is a moderately coupled electron-phonon superconductor with strong spin-orbit coupling effects on its physical properties. 

\section{Acknowledgments}
This work was partly supported by the National Science Center (Poland), grant No. 2017/26/E/ST3/00119, and by the AGH-UST statutory tasks No. 11.11.220.01/5 within subsidy of the Ministry of Science and Higher Education.
\bibliography{references}

\begin{thebibliography}{53}%
\makeatletter
\providecommand \@ifxundefined [1]{%
 \@ifx{#1\undefined}
}%
\providecommand \@ifnum [1]{%
 \ifnum #1\expandafter \@firstoftwo
 \else \expandafter \@secondoftwo
 \fi
}%
\providecommand \@ifx [1]{%
 \ifx #1\expandafter \@firstoftwo
 \else \expandafter \@secondoftwo
 \fi
}%
\providecommand \natexlab [1]{#1}%
\providecommand \enquote  [1]{``#1''}%
\providecommand \bibnamefont  [1]{#1}%
\providecommand \bibfnamefont [1]{#1}%
\providecommand \citenamefont [1]{#1}%
\providecommand \href@noop [0]{\@secondoftwo}%
\providecommand \href [0]{\begingroup \@sanitize@url \@href}%
\providecommand \@href[1]{\@@startlink{#1}\@@href}%
\providecommand \@@href[1]{\endgroup#1\@@endlink}%
\providecommand \@sanitize@url [0]{\catcode `\\12\catcode `\$12\catcode
  `\&12\catcode `\#12\catcode `\^12\catcode `\_12\catcode `\%12\relax}%
\providecommand \@@startlink[1]{}%
\providecommand \@@endlink[0]{}%
\providecommand \url  [0]{\begingroup\@sanitize@url \@url }%
\providecommand \@url [1]{\endgroup\@href {#1}{\urlprefix }}%
\providecommand \urlprefix  [0]{URL }%
\providecommand \Eprint [0]{\href }%
\providecommand \doibase [0]{http://dx.doi.org/}%
\providecommand \selectlanguage [0]{\@gobble}%
\providecommand \bibinfo  [0]{\@secondoftwo}%
\providecommand \bibfield  [0]{\@secondoftwo}%
\providecommand \translation [1]{[#1]}%
\providecommand \BibitemOpen [0]{}%
\providecommand \bibitemStop [0]{}%
\providecommand \bibitemNoStop [0]{.\EOS\space}%
\providecommand \EOS [0]{\spacefactor3000\relax}%
\providecommand \BibitemShut  [1]{\csname bibitem#1\endcsname}%
\let\auto@bib@innerbib\@empty
\bibitem [{\citenamefont {Jones}(1934)}]{Bi-jones}%
  \BibitemOpen
  \bibfield  {author} {\bibinfo {author} {\bibfnamefont {H.}~\bibnamefont
  {Jones}},\ }\bibfield  {title} {\enquote {\bibinfo {title} {{Applications of
  the {B}loch Theory to the Study of Alloys and of the Properties of
  Bismuth}},}\ }\href@noop {} {\bibfield  {journal} {\bibinfo  {journal}
  {Proceedings of the Royal Society of London A}\ }\textbf {\bibinfo {volume}
  {147}},\ \bibinfo {pages} {396--417} (\bibinfo {year} {1934})}\BibitemShut
  {NoStop}%
\bibitem [{\citenamefont {Fuseya}\ \emph {et~al.}(2015)\citenamefont {Fuseya},
  \citenamefont {Ogata},\ and\ \citenamefont {Fukuyama}}]{Bi-szescian}%
  \BibitemOpen
  \bibfield  {author} {\bibinfo {author} {\bibfnamefont {Y.}~\bibnamefont
  {Fuseya}}, \bibinfo {author} {\bibfnamefont {M.}~\bibnamefont {Ogata}}, \
  and\ \bibinfo {author} {\bibfnamefont {H.}~\bibnamefont {Fukuyama}},\
  }\bibfield  {title} {\enquote {\bibinfo {title} {{Transport Properties and
  Diamagnetism of {D}irac Electrons in Bismuth}},}\ }\href@noop {} {\bibfield
  {journal} {\bibinfo  {journal} {Journal of the Physical Society of Japan}\
  }\textbf {\bibinfo {volume} {84}},\ \bibinfo {pages} {012001} (\bibinfo
  {year} {2015})}\BibitemShut {NoStop}%
\bibitem [{\citenamefont {{{\'E}del'Man}}\ and\ \citenamefont {{Kha{\v
  i}kin}}(1966)}]{bi-fs}%
  \BibitemOpen
  \bibfield  {author} {\bibinfo {author} {\bibfnamefont {V.~S.}\ \bibnamefont
  {{{\'E}del'Man}}}\ and\ \bibinfo {author} {\bibfnamefont {M.~S.}\
  \bibnamefont {{Kha{\v i}kin}}},\ }\bibfield  {title} {\enquote {\bibinfo
  {title} {{Investigation of the Fermi Surface in Bismuth by Means of Cyclotron
  Resonance}},}\ }\href@noop {} {\bibfield  {journal} {\bibinfo  {journal}
  {Soviet Journal of Experimental and Theoretical Physics}\ }\textbf {\bibinfo
  {volume} {22}},\ \bibinfo {pages} {77} (\bibinfo {year} {1966})}\BibitemShut
  {NoStop}%
\bibitem [{\citenamefont {Jin}\ \emph {et~al.}(2015)\citenamefont {Jin},
  \citenamefont {Wiendlocha},\ and\ \citenamefont {Heremans}}]{bi-in}%
  \BibitemOpen
  \bibfield  {author} {\bibinfo {author} {\bibfnamefont {Hyungyu}\ \bibnamefont
  {Jin}}, \bibinfo {author} {\bibfnamefont {Bartlomiej}\ \bibnamefont
  {Wiendlocha}}, \ and\ \bibinfo {author} {\bibfnamefont {Joseph~P.}\
  \bibnamefont {Heremans}},\ }\bibfield  {title} {\enquote {\bibinfo {title}
  {{{P-type doping of elemental bismuth with indium{,} gallium and tin: a novel
  doping mechanism in solids}}},}\ }\href {\doibase 10.1039/C5EE01309G}
  {\bibfield  {journal} {\bibinfo  {journal} {Energy Environ. Sci.}\ }\textbf
  {\bibinfo {volume} {8}},\ \bibinfo {pages} {2027--2040} (\bibinfo {year}
  {2015})}\BibitemShut {NoStop}%
\bibitem [{\citenamefont {{Zhu}}\ \emph {et~al.}(2012)\citenamefont {{Zhu}},
  \citenamefont {{Collaudin}}, \citenamefont {{Fauqu{\'e}}}, \citenamefont
  {{Kang}},\ and\ \citenamefont {{Behnia}}}]{Bi-valleytronic}%
  \BibitemOpen
  \bibfield  {author} {\bibinfo {author} {\bibfnamefont {Z.}~\bibnamefont
  {{Zhu}}}, \bibinfo {author} {\bibfnamefont {A.}~\bibnamefont {{Collaudin}}},
  \bibinfo {author} {\bibfnamefont {B.}~\bibnamefont {{Fauqu{\'e}}}}, \bibinfo
  {author} {\bibfnamefont {W.}~\bibnamefont {{Kang}}}, \ and\ \bibinfo {author}
  {\bibfnamefont {K.}~\bibnamefont {{Behnia}}},\ }\bibfield  {title} {\enquote
  {\bibinfo {title} {{Field-induced polarization of Dirac valleys in
  Bismuth}},}\ }\href {\doibase 10.1038/nphys2111} {\bibfield  {journal}
  {\bibinfo  {journal} {Nature Physics}\ }\textbf {\bibinfo {volume} {8}},\
  \bibinfo {pages} {89--94} (\bibinfo {year} {2012})},\ \Eprint
  {http://arxiv.org/abs/1109.2774} {arXiv:1109.2774 [cond-mat.str-el]}
  \BibitemShut {NoStop}%
\bibitem [{\citenamefont {Buckel}\ and\ \citenamefont
  {Hilsch}(1954)}]{Buckel1954}%
  \BibitemOpen
  \bibfield  {author} {\bibinfo {author} {\bibfnamefont {W.}~\bibnamefont
  {Buckel}}\ and\ \bibinfo {author} {\bibfnamefont {R.}~\bibnamefont
  {Hilsch}},\ }\bibfield  {title} {\enquote {\bibinfo {title} {{Einflu{\ss} der
  Kondensation bei tiefen Temperaturen auf den elektrischen Widerstand und die
  Supraleitung f{\"u}r verschiedene Metalle}},}\ }\href {\doibase
  10.1007/BF01337903} {\bibfield  {journal} {\bibinfo  {journal} {Zeitschrift
  f{\"u}r Physik}\ }\textbf {\bibinfo {volume} {138}},\ \bibinfo {pages}
  {109--120} (\bibinfo {year} {1954})}\BibitemShut {NoStop}%
\bibitem [{\citenamefont {Mata-Pinz{\'o}n}\ \emph {et~al.}(2016)\citenamefont
  {Mata-Pinz{\'o}n}, \citenamefont {A.}, \citenamefont {M.},\ and\
  \citenamefont {Alexander}}]{Bi-amorphous-sc}%
  \BibitemOpen
  \bibfield  {author} {\bibinfo {author} {\bibfnamefont {Zaahel}\ \bibnamefont
  {Mata-Pinz{\'o}n}}, \bibinfo {author} {\bibfnamefont {Valladares~Ariel}\
  \bibnamefont {A.}}, \bibinfo {author} {\bibfnamefont {Valladares~Renela}\
  \bibnamefont {M.}}, \ and\ \bibinfo {author} {\bibfnamefont {Valladares}\
  \bibnamefont {Alexander}},\ }\bibfield  {title} {\enquote {\bibinfo {title}
  {{{Superconductivity in Bismuth. A New Look at an Old Problem}}},}\ }\href
  {\doibase 10.1371/journal.pone.0147645} {\bibfield  {journal} {\bibinfo
  {journal} {PLOS ONE}\ }\textbf {\bibinfo {volume} {11}},\ \bibinfo {pages}
  {1--20} (\bibinfo {year} {2016})}\BibitemShut {NoStop}%
\bibitem [{\citenamefont {Prakash}\ \emph {et~al.}(2017)\citenamefont
  {Prakash}, \citenamefont {Kumar}, \citenamefont {Thamizhavel},\ and\
  \citenamefont {Ramakrishnan}}]{Bi-sc}%
  \BibitemOpen
  \bibfield  {author} {\bibinfo {author} {\bibfnamefont {Om}~\bibnamefont
  {Prakash}}, \bibinfo {author} {\bibfnamefont {Anil}\ \bibnamefont {Kumar}},
  \bibinfo {author} {\bibfnamefont {A}~\bibnamefont {Thamizhavel}}, \ and\
  \bibinfo {author} {\bibfnamefont {S}~\bibnamefont {Ramakrishnan}},\
  }\bibfield  {title} {\enquote {\bibinfo {title} {{Evidence for bulk
  superconductivity in pure bismuth single crystals at ambient pressure}},}\
  }\href@noop {} {\bibfield  {journal} {\bibinfo  {journal} {Science}\ }\textbf
  {\bibinfo {volume} {355}},\ \bibinfo {pages} {52--55} (\bibinfo {year}
  {2017})}\BibitemShut {NoStop}%
\bibitem [{\citenamefont {Shamrai}(2013)}]{bi-high-tc-sc}%
  \BibitemOpen
  \bibfield  {author} {\bibinfo {author} {\bibfnamefont {V.~F.}\ \bibnamefont
  {Shamrai}},\ }\bibfield  {title} {\enquote {\bibinfo {title} {{Crystal
  Structures and Superconductivity of Bismuth High Temperature Superconductors
  (Review)}},}\ }\href@noop {} {\bibfield  {journal} {\bibinfo  {journal}
  {Inorganic Materials: Applied Research}\ }\textbf {\bibinfo {volume} {4}},\
  \bibinfo {pages} {273} (\bibinfo {year} {2013})}\BibitemShut {NoStop}%
\bibitem [{\citenamefont {Shao}\ \emph {et~al.}(2016)\citenamefont {Shao},
  \citenamefont {Luo}, \citenamefont {Lu}, \citenamefont {Hu}, \citenamefont
  {Zhu}, \citenamefont {Song}, \citenamefont {Zhu},\ and\ \citenamefont
  {Sun}}]{abi3}%
  \BibitemOpen
  \bibfield  {author} {\bibinfo {author} {\bibfnamefont {D.~F.}\ \bibnamefont
  {Shao}}, \bibinfo {author} {\bibfnamefont {X.}~\bibnamefont {Luo}}, \bibinfo
  {author} {\bibfnamefont {W.~J.}\ \bibnamefont {Lu}}, \bibinfo {author}
  {\bibfnamefont {L.}~\bibnamefont {Hu}}, \bibinfo {author} {\bibfnamefont
  {X.~D.}\ \bibnamefont {Zhu}}, \bibinfo {author} {\bibfnamefont {W.~H.}\
  \bibnamefont {Song}}, \bibinfo {author} {\bibfnamefont {X.~B.}\ \bibnamefont
  {Zhu}}, \ and\ \bibinfo {author} {\bibfnamefont {Y.~P.}\ \bibnamefont
  {Sun}},\ }\bibfield  {title} {\enquote {\bibinfo {title} {{Spin-orbit
  coupling enhanced superconductivity in {Bi}-rich compounds {ABi}${}_3$ (A =
  Sr and Ba)}},}\ }\href@noop {} {\bibfield  {journal} {\bibinfo  {journal}
  {Scientific Reports}\ }\textbf {\bibinfo {volume} {6}},\ \bibinfo {pages}
  {21484} (\bibinfo {year} {2016})}\BibitemShut {NoStop}%
\bibitem [{\citenamefont {Matthias}\ and\ \citenamefont
  {Hulm}(1952)}]{abi3-exp-1st}%
  \BibitemOpen
  \bibfield  {author} {\bibinfo {author} {\bibfnamefont {B.~T.}\ \bibnamefont
  {Matthias}}\ and\ \bibinfo {author} {\bibfnamefont {J.~K.}\ \bibnamefont
  {Hulm}},\ }\bibfield  {title} {\enquote {\bibinfo {title} {{A Search for New
  Superconducting Compounds}},}\ }\href {\doibase 10.1103/PhysRev.87.799}
  {\bibfield  {journal} {\bibinfo  {journal} {Phys. Rev.}\ }\textbf {\bibinfo
  {volume} {87}},\ \bibinfo {pages} {799--806} (\bibinfo {year}
  {1952})}\BibitemShut {NoStop}%
\bibitem [{\citenamefont {Gati}\ \emph {et~al.}(2018)\citenamefont {Gati},
  \citenamefont {Xiang}, \citenamefont {Wang}, \citenamefont {Manni},
  \citenamefont {Canfield},\ and\ \citenamefont {Budko}}]{nibi3}%
  \BibitemOpen
  \bibfield  {author} {\bibinfo {author} {\bibfnamefont {Elena}\ \bibnamefont
  {Gati}}, \bibinfo {author} {\bibfnamefont {Li}~\bibnamefont {Xiang}},
  \bibinfo {author} {\bibfnamefont {Lin-Lin}\ \bibnamefont {Wang}}, \bibinfo
  {author} {\bibfnamefont {Soham}\ \bibnamefont {Manni}}, \bibinfo {author}
  {\bibfnamefont {Paul~C}\ \bibnamefont {Canfield}}, \ and\ \bibinfo {author}
  {\bibfnamefont {Sergey~L}\ \bibnamefont {Budko}},\ }\bibfield  {title}
  {\enquote {\bibinfo {title} {{Effect of pressure on the physical properties
  of the superconductor NiBi${}_3$}},}\ }\href@noop {} {\bibfield  {journal}
  {\bibinfo  {journal} {Journal of Physics: Condensed Matter}\ }\textbf
  {\bibinfo {volume} {31}},\ \bibinfo {pages} {035701} (\bibinfo {year}
  {2018})}\BibitemShut {NoStop}%
\bibitem [{\citenamefont {Kinjo}\ \emph {et~al.}(2016)\citenamefont {Kinjo},
  \citenamefont {Kajino}, \citenamefont {Nishio}, \citenamefont {Kawashima},
  \citenamefont {Yanagi}, \citenamefont {Hase}, \citenamefont {Yanagisawa},
  \citenamefont {Ishida}, \citenamefont {Kito}, \citenamefont {Takeshita} \emph
  {et~al.}}]{labi3}%
  \BibitemOpen
  \bibfield  {author} {\bibinfo {author} {\bibfnamefont {Tatsuya}\ \bibnamefont
  {Kinjo}}, \bibinfo {author} {\bibfnamefont {Saori}\ \bibnamefont {Kajino}},
  \bibinfo {author} {\bibfnamefont {Taichiro}\ \bibnamefont {Nishio}}, \bibinfo
  {author} {\bibfnamefont {Kenji}\ \bibnamefont {Kawashima}}, \bibinfo {author}
  {\bibfnamefont {Yousuke}\ \bibnamefont {Yanagi}}, \bibinfo {author}
  {\bibfnamefont {Izumi}\ \bibnamefont {Hase}}, \bibinfo {author}
  {\bibfnamefont {Takashi}\ \bibnamefont {Yanagisawa}}, \bibinfo {author}
  {\bibfnamefont {Shigeyuki}\ \bibnamefont {Ishida}}, \bibinfo {author}
  {\bibfnamefont {Hijiri}\ \bibnamefont {Kito}}, \bibinfo {author}
  {\bibfnamefont {Nao}\ \bibnamefont {Takeshita}},  \emph {et~al.},\ }\bibfield
   {title} {\enquote {\bibinfo {title} {{Superconductivity in LaBi${}_3$ with
  AuCu${}_3$-type structure}},}\ }\href@noop {} {\bibfield  {journal} {\bibinfo
   {journal} {Superconductor Science and Technology}\ }\textbf {\bibinfo
  {volume} {29}},\ \bibinfo {pages} {03LT02} (\bibinfo {year}
  {2016})}\BibitemShut {NoStop}%
\bibitem [{\citenamefont {Tenc{\'e}}\ \emph {et~al.}(2014)\citenamefont
  {Tenc{\'e}}, \citenamefont {Janson}, \citenamefont {Krellner}, \citenamefont
  {Rosner}, \citenamefont {Schwarz}, \citenamefont {Grin},\ and\ \citenamefont
  {Steglich}}]{cobi3}%
  \BibitemOpen
  \bibfield  {author} {\bibinfo {author} {\bibfnamefont {Sophie}\ \bibnamefont
  {Tenc{\'e}}}, \bibinfo {author} {\bibfnamefont {Oleg}\ \bibnamefont
  {Janson}}, \bibinfo {author} {\bibfnamefont {Cornelius}\ \bibnamefont
  {Krellner}}, \bibinfo {author} {\bibfnamefont {Helge}\ \bibnamefont
  {Rosner}}, \bibinfo {author} {\bibfnamefont {Ulrich}\ \bibnamefont
  {Schwarz}}, \bibinfo {author} {\bibfnamefont {Y}~\bibnamefont {Grin}}, \ and\
  \bibinfo {author} {\bibfnamefont {F}~\bibnamefont {Steglich}},\ }\bibfield
  {title} {\enquote {\bibinfo {title} {{CoBi${}_3$ - the first binary compound
  of cobalt with bismuth: high-pressure synthesis and superconductivity}},}\
  }\href@noop {} {\bibfield  {journal} {\bibinfo  {journal} {Journal of
  Physics: Condensed Matter}\ }\textbf {\bibinfo {volume} {26}},\ \bibinfo
  {pages} {395701} (\bibinfo {year} {2014})}\BibitemShut {NoStop}%
\bibitem [{\citenamefont {Sambongi}(1971)}]{libi}%
  \BibitemOpen
  \bibfield  {author} {\bibinfo {author} {\bibfnamefont {Takashi}\ \bibnamefont
  {Sambongi}},\ }\bibfield  {title} {\enquote {\bibinfo {title}
  {{Superconductivity of LiBi}},}\ }\href@noop {} {\bibfield  {journal}
  {\bibinfo  {journal} {Journal of the Physical Society of Japan}\ }\textbf
  {\bibinfo {volume} {30}},\ \bibinfo {pages} {294--294} (\bibinfo {year}
  {1971})}\BibitemShut {NoStop}%
\bibitem [{\citenamefont {Kushwaha}\ \emph {et~al.}(2014)\citenamefont
  {Kushwaha}, \citenamefont {Krizan}, \citenamefont {Xiong}, \citenamefont
  {Klimczuk}, \citenamefont {Gibson}, \citenamefont {Liang}, \citenamefont
  {Ong},\ and\ \citenamefont {Cava}}]{nabi}%
  \BibitemOpen
  \bibfield  {author} {\bibinfo {author} {\bibfnamefont {SK}~\bibnamefont
  {Kushwaha}}, \bibinfo {author} {\bibfnamefont {JW}~\bibnamefont {Krizan}},
  \bibinfo {author} {\bibfnamefont {Jun}\ \bibnamefont {Xiong}}, \bibinfo
  {author} {\bibfnamefont {Tomasz}\ \bibnamefont {Klimczuk}}, \bibinfo {author}
  {\bibfnamefont {QD}~\bibnamefont {Gibson}}, \bibinfo {author} {\bibfnamefont
  {Tian}\ \bibnamefont {Liang}}, \bibinfo {author} {\bibfnamefont
  {NP}~\bibnamefont {Ong}}, \ and\ \bibinfo {author} {\bibfnamefont
  {RJ}~\bibnamefont {Cava}},\ }\bibfield  {title} {\enquote {\bibinfo {title}
  {{Superconducting properties and electronic structure of NaBi}},}\
  }\href@noop {} {\bibfield  {journal} {\bibinfo  {journal} {Journal of
  Physics: Condensed Matter}\ }\textbf {\bibinfo {volume} {26}},\ \bibinfo
  {pages} {212201} (\bibinfo {year} {2014})}\BibitemShut {NoStop}%
\bibitem [{\citenamefont {Roberts}(1976)}]{abi2-tc}%
  \BibitemOpen
  \bibfield  {author} {\bibinfo {author} {\bibfnamefont {B.~W.}\ \bibnamefont
  {Roberts}},\ }\bibfield  {title} {\enquote {\bibinfo {title} {{Survey of
  superconductive materials and critical evaluation of selected properties}},}\
  }\href {\doibase 10.1063/1.555540} {\bibfield  {journal} {\bibinfo  {journal}
  {Journal of Physical and Chemical Reference Data}\ }\textbf {\bibinfo
  {volume} {5}},\ \bibinfo {pages} {581--822} (\bibinfo {year}
  {1976})}\BibitemShut {NoStop}%
\bibitem [{\citenamefont {Winiarski}\ \emph {et~al.}(2016)\citenamefont
  {Winiarski}, \citenamefont {Wiendlocha}, \citenamefont {Golab}, \citenamefont
  {Kushwaha}, \citenamefont {Wiśniewski}, \citenamefont {Kaczorowski},
  \citenamefont {Thompson}, \citenamefont {Cava},\ and\ \citenamefont
  {Klimczuk}}]{cabi2}%
  \BibitemOpen
  \bibfield  {author} {\bibinfo {author} {\bibfnamefont {M.J.}\ \bibnamefont
  {Winiarski}}, \bibinfo {author} {\bibfnamefont {B.}~\bibnamefont
  {Wiendlocha}}, \bibinfo {author} {\bibfnamefont {S.}~\bibnamefont {Golab}},
  \bibinfo {author} {\bibfnamefont {S.~K.}\ \bibnamefont {Kushwaha}}, \bibinfo
  {author} {\bibfnamefont {P.}~\bibnamefont {Wiśniewski}}, \bibinfo {author}
  {\bibfnamefont {D.}~\bibnamefont {Kaczorowski}}, \bibinfo {author}
  {\bibfnamefont {J.~D.}\ \bibnamefont {Thompson}}, \bibinfo {author}
  {\bibfnamefont {R.~J.}\ \bibnamefont {Cava}}, \ and\ \bibinfo {author}
  {\bibfnamefont {T.}~\bibnamefont {Klimczuk}},\ }\bibfield  {title} {\enquote
  {\bibinfo {title} {{Superconductivity in {CaBi}${}_2$}},}\ }\href@noop {}
  {\bibfield  {journal} {\bibinfo  {journal} {Physical Chemistry Chemical
  Physics}\ }\textbf {\bibinfo {volume} {18(31)}},\ \bibinfo {pages} {21737}
  (\bibinfo {year} {2016})}\BibitemShut {NoStop}%
\bibitem [{\citenamefont {Hasan}\ and\ \citenamefont {Kane}(2010)}]{coloqium}%
  \BibitemOpen
  \bibfield  {author} {\bibinfo {author} {\bibfnamefont {M~Zahid}\ \bibnamefont
  {Hasan}}\ and\ \bibinfo {author} {\bibfnamefont {Charles~L}\ \bibnamefont
  {Kane}},\ }\bibfield  {title} {\enquote {\bibinfo {title} {Colloquium:
  topological insulators},}\ }\href@noop {} {\bibfield  {journal} {\bibinfo
  {journal} {Reviews of modern physics}\ }\textbf {\bibinfo {volume} {82}},\
  \bibinfo {pages} {3045} (\bibinfo {year} {2010})}\BibitemShut {NoStop}%
\bibitem [{\citenamefont {{Heremans}}\ \emph {et~al.}(2017)\citenamefont
  {{Heremans}}, \citenamefont {{Cava}},\ and\ \citenamefont
  {{Samarth}}}]{heremans_rev}%
  \BibitemOpen
  \bibfield  {author} {\bibinfo {author} {\bibfnamefont {J.~P.}\ \bibnamefont
  {{Heremans}}}, \bibinfo {author} {\bibfnamefont {R.~J.}\ \bibnamefont
  {{Cava}}}, \ and\ \bibinfo {author} {\bibfnamefont {N.}~\bibnamefont
  {{Samarth}}},\ }\bibfield  {title} {\enquote {\bibinfo {title} {{Tetradymites
  as thermoelectrics and topological insulators}},}\ }\href {\doibase
  10.1038/natrevmats.2017.49} {\bibfield  {journal} {\bibinfo  {journal}
  {Nature Reviews Materials}\ }\textbf {\bibinfo {volume} {2}},\ \bibinfo
  {pages} {17049} (\bibinfo {year} {2017})}\BibitemShut {NoStop}%
\bibitem [{\citenamefont {Du}\ \emph {et~al.}(2017)\citenamefont {Du},
  \citenamefont {Shao}, \citenamefont {Yang}, \citenamefont {du}, \citenamefont
  {Fang}, \citenamefont {Zhang}, \citenamefont {Wang}, \citenamefont {Ran},
  \citenamefont {Wen}, \citenamefont {Yang}, \citenamefont {Zhang},\ and\
  \citenamefont {Wen}}]{srbise}%
  \BibitemOpen
  \bibfield  {author} {\bibinfo {author} {\bibfnamefont {Guan}\ \bibnamefont
  {Du}}, \bibinfo {author} {\bibfnamefont {Jifeng}\ \bibnamefont {Shao}},
  \bibinfo {author} {\bibfnamefont {Xiong}\ \bibnamefont {Yang}}, \bibinfo
  {author} {\bibfnamefont {Zengyi}\ \bibnamefont {du}}, \bibinfo {author}
  {\bibfnamefont {Delong}\ \bibnamefont {Fang}}, \bibinfo {author}
  {\bibfnamefont {Changjing}\ \bibnamefont {Zhang}}, \bibinfo {author}
  {\bibfnamefont {Jinghui}\ \bibnamefont {Wang}}, \bibinfo {author}
  {\bibfnamefont {Kejing}\ \bibnamefont {Ran}}, \bibinfo {author}
  {\bibfnamefont {Jinsheng}\ \bibnamefont {Wen}}, \bibinfo {author}
  {\bibfnamefont {Huan}\ \bibnamefont {Yang}}, \bibinfo {author} {\bibfnamefont
  {Yuheng}\ \bibnamefont {Zhang}}, \ and\ \bibinfo {author} {\bibfnamefont
  {Hai-Hu}\ \bibnamefont {Wen}},\ }\bibfield  {title} {\enquote {\bibinfo
  {title} {{Drive the Dirac electrons into Cooper pairs in
  Sr${}_x$Bi${}_2$Se${}_3$}},}\ }\href@noop {} {\bibfield  {journal} {\bibinfo
  {journal} {Nature communications}\ }\textbf {\bibinfo {volume} {8}},\
  \bibinfo {pages} {14466} (\bibinfo {year} {2017})}\BibitemShut {NoStop}%
\bibitem [{\citenamefont {Li}\ \emph {et~al.}(2015)\citenamefont {Li},
  \citenamefont {Xie}, \citenamefont {Cheng}, \citenamefont {Li}, \citenamefont
  {Li},\ and\ \citenamefont {Chen}}]{m3bi2}%
  \BibitemOpen
  \bibfield  {author} {\bibinfo {author} {\bibfnamefont {Ronghan}\ \bibnamefont
  {Li}}, \bibinfo {author} {\bibfnamefont {Qing}\ \bibnamefont {Xie}}, \bibinfo
  {author} {\bibfnamefont {Xiyue}\ \bibnamefont {Cheng}}, \bibinfo {author}
  {\bibfnamefont {Dianzhong}\ \bibnamefont {Li}}, \bibinfo {author}
  {\bibfnamefont {Yiyi}\ \bibnamefont {Li}}, \ and\ \bibinfo {author}
  {\bibfnamefont {Xing-Qiu}\ \bibnamefont {Chen}},\ }\bibfield  {title}
  {\enquote {\bibinfo {title} {{First-principles study of the large-gap
  three-dimensional topological insulators M${}_3$Bi${}_2$ (M= Ca, Sr, Ba)}},}\
  }\href@noop {} {\bibfield  {journal} {\bibinfo  {journal} {Physical Review
  B}\ }\textbf {\bibinfo {volume} {92}},\ \bibinfo {pages} {205130} (\bibinfo
  {year} {2015})}\BibitemShut {NoStop}%
\bibitem [{\citenamefont {Nourbakhsh}\ and\ \citenamefont {Vaez}(2016)}]{ThXY}%
  \BibitemOpen
  \bibfield  {author} {\bibinfo {author} {\bibfnamefont {Zahra}\ \bibnamefont
  {Nourbakhsh}}\ and\ \bibinfo {author} {\bibfnamefont {Aminollah}\
  \bibnamefont {Vaez}},\ }\bibfield  {title} {\enquote {\bibinfo {title}
  {{Electronic properties and topological phases of Th XY ( X = Pb, Au, Pt and
  Y = Sb, Bi, Sn) compounds}},}\ }\bibfield  {booktitle} {\emph {\bibinfo
  {booktitle} {{Chinese Physics B}}},\ }\href@noop {} {\ \textbf {\bibinfo
  {volume} {25}},\ \bibinfo {pages} {037101} (\bibinfo {year}
  {2016})}\BibitemShut {NoStop}%
\bibitem [{\citenamefont {Wang}\ and\ \citenamefont {Wei}(2016)}]{hfirbi}%
  \BibitemOpen
  \bibfield  {author} {\bibinfo {author} {\bibfnamefont {Guangtao}\
  \bibnamefont {Wang}}\ and\ \bibinfo {author} {\bibfnamefont {JunHong}\
  \bibnamefont {Wei}},\ }\bibfield  {title} {\enquote {\bibinfo {title}
  {{Topological phase transition in half-Heusler compounds HfIrX (X=As, Sb,
  Bi)}},}\ }\href {\doibase https://doi.org/10.1016/j.commatsci.2016.08.005}
  {\bibfield  {journal} {\bibinfo  {journal} {Computational Materials Science}\
  }\textbf {\bibinfo {volume} {124}},\ \bibinfo {pages} {311 -- 315} (\bibinfo
  {year} {2016})}\BibitemShut {NoStop}%
\bibitem [{\citenamefont {Huang}\ and\ \citenamefont {Duan}(2016)}]{bii-1d}%
  \BibitemOpen
  \bibfield  {author} {\bibinfo {author} {\bibfnamefont {Huaqing}\ \bibnamefont
  {Huang}}\ and\ \bibinfo {author} {\bibfnamefont {Wenhui}\ \bibnamefont
  {Duan}},\ }\bibfield  {title} {\enquote {\bibinfo {title} {{Topological
  insulators: Quasi-1D topological insulators}},}\ }\bibfield  {booktitle}
  {\emph {\bibinfo {booktitle} {{Nature Materials}}},\ }\href@noop {} {\
  \textbf {\bibinfo {volume} {15}},\ \bibinfo {pages} {129--130} (\bibinfo
  {year} {2016})}\BibitemShut {NoStop}%
\bibitem [{\citenamefont {Koelling}\ and\ \citenamefont
  {N~Harmon}(1977)}]{scalar}%
  \BibitemOpen
  \bibfield  {author} {\bibinfo {author} {\bibfnamefont {Dale}\ \bibnamefont
  {Koelling}}\ and\ \bibinfo {author} {\bibfnamefont {B}~\bibnamefont
  {N~Harmon}},\ }\bibfield  {title} {\enquote {\bibinfo {title} {{A technique
  for relativistic spin-polarised calculations}},}\ }\bibfield  {booktitle}
  {\emph {\bibinfo {booktitle} {{Journal of Physics C: Solid State Physics}}},\
  }\href@noop {} {\ \textbf {\bibinfo {volume} {10}},\ \bibinfo {pages} {3107}
  (\bibinfo {year} {1977})}\BibitemShut {NoStop}%
\bibitem [{Note1()}]{Note1}%
  \BibitemOpen
  \bibinfo {note} {See Fig. S1 in Supplemental Material for the relation
  between the conventional and primitive cells.}\BibitemShut {Stop}%
\bibitem [{\citenamefont {Giannozzi}\ \emph {et~al.}(2009)\citenamefont
  {Giannozzi}, \citenamefont {Baroni}, \citenamefont {Bonini}, \citenamefont
  {Calandra}, \citenamefont {Car}, \citenamefont {Cavazzoni}, \citenamefont
  {Ceresoli}, \citenamefont {Chiarotti}, \citenamefont {Cococcioni},
  \citenamefont {Dabo}, \citenamefont {{Dal Corso}}, \citenamefont
  {de~Gironcoli}, \citenamefont {Fabris}, \citenamefont {Fratesi},
  \citenamefont {Gebauer}, \citenamefont {Gerstmann}, \citenamefont
  {Gougoussis}, \citenamefont {Kokalj}, \citenamefont {Lazzeri}, \citenamefont
  {Martin-Samos}, \citenamefont {Marzari}, \citenamefont {Mauri}, \citenamefont
  {Mazzarello}, \citenamefont {Paolini}, \citenamefont {Pasquarello},
  \citenamefont {Paulatto}, \citenamefont {Sbraccia}, \citenamefont {Scandolo},
  \citenamefont {Sclauzero}, \citenamefont {Seitsonen}, \citenamefont
  {Smogunov}, \citenamefont {Umari},\ and\ \citenamefont
  {Wentzcovitch}}]{QE-2009}%
  \BibitemOpen
  \bibfield  {author} {\bibinfo {author} {\bibfnamefont {Paolo}\ \bibnamefont
  {Giannozzi}}, \bibinfo {author} {\bibfnamefont {Stefano}\ \bibnamefont
  {Baroni}}, \bibinfo {author} {\bibfnamefont {Nicola}\ \bibnamefont {Bonini}},
  \bibinfo {author} {\bibfnamefont {Matteo}\ \bibnamefont {Calandra}}, \bibinfo
  {author} {\bibfnamefont {Roberto}\ \bibnamefont {Car}}, \bibinfo {author}
  {\bibfnamefont {Carlo}\ \bibnamefont {Cavazzoni}}, \bibinfo {author}
  {\bibfnamefont {Davide}\ \bibnamefont {Ceresoli}}, \bibinfo {author}
  {\bibfnamefont {Guido~L}\ \bibnamefont {Chiarotti}}, \bibinfo {author}
  {\bibfnamefont {Matteo}\ \bibnamefont {Cococcioni}}, \bibinfo {author}
  {\bibfnamefont {Ismaila}\ \bibnamefont {Dabo}}, \bibinfo {author}
  {\bibfnamefont {Andrea}\ \bibnamefont {{Dal Corso}}}, \bibinfo {author}
  {\bibfnamefont {Stefano}\ \bibnamefont {de~Gironcoli}}, \bibinfo {author}
  {\bibfnamefont {Stefano}\ \bibnamefont {Fabris}}, \bibinfo {author}
  {\bibfnamefont {Guido}\ \bibnamefont {Fratesi}}, \bibinfo {author}
  {\bibfnamefont {Ralph}\ \bibnamefont {Gebauer}}, \bibinfo {author}
  {\bibfnamefont {Uwe}\ \bibnamefont {Gerstmann}}, \bibinfo {author}
  {\bibfnamefont {Christos}\ \bibnamefont {Gougoussis}}, \bibinfo {author}
  {\bibfnamefont {Anton}\ \bibnamefont {Kokalj}}, \bibinfo {author}
  {\bibfnamefont {Michele}\ \bibnamefont {Lazzeri}}, \bibinfo {author}
  {\bibfnamefont {Layla}\ \bibnamefont {Martin-Samos}}, \bibinfo {author}
  {\bibfnamefont {Nicola}\ \bibnamefont {Marzari}}, \bibinfo {author}
  {\bibfnamefont {Francesco}\ \bibnamefont {Mauri}}, \bibinfo {author}
  {\bibfnamefont {Riccardo}\ \bibnamefont {Mazzarello}}, \bibinfo {author}
  {\bibfnamefont {Stefano}\ \bibnamefont {Paolini}}, \bibinfo {author}
  {\bibfnamefont {Alfredo}\ \bibnamefont {Pasquarello}}, \bibinfo {author}
  {\bibfnamefont {Lorenzo}\ \bibnamefont {Paulatto}}, \bibinfo {author}
  {\bibfnamefont {Carlo}\ \bibnamefont {Sbraccia}}, \bibinfo {author}
  {\bibfnamefont {Sandro}\ \bibnamefont {Scandolo}}, \bibinfo {author}
  {\bibfnamefont {Gabriele}\ \bibnamefont {Sclauzero}}, \bibinfo {author}
  {\bibfnamefont {Ari~P}\ \bibnamefont {Seitsonen}}, \bibinfo {author}
  {\bibfnamefont {Alexander}\ \bibnamefont {Smogunov}}, \bibinfo {author}
  {\bibfnamefont {Paolo}\ \bibnamefont {Umari}}, \ and\ \bibinfo {author}
  {\bibfnamefont {Renata~M}\ \bibnamefont {Wentzcovitch}},\ }\bibfield  {title}
  {\enquote {\bibinfo {title} {{QUANTUM ESPRESSO: a modular and open-source
  software project for quantum simulations of materials}},}\ }\href
  {http://www.quantum-espresso.org} {\bibfield  {journal} {\bibinfo  {journal}
  {Journal of Physics: Condensed Matter}\ }\textbf {\bibinfo {volume} {21}},\
  \bibinfo {pages} {395502 (19pp)} (\bibinfo {year} {2009})}\BibitemShut
  {NoStop}%
\bibitem [{\citenamefont {Giannozzi}\ \emph {et~al.}(2017)\citenamefont
  {Giannozzi}, \citenamefont {Andreussi}, \citenamefont {Brumme}, \citenamefont
  {Bunau}, \citenamefont {Nardelli}, \citenamefont {Calandra}, \citenamefont
  {Car}, \citenamefont {Cavazzoni}, \citenamefont {Ceresoli}, \citenamefont
  {Cococcioni}, \citenamefont {Colonna}, \citenamefont {Carnimeo},
  \citenamefont {Corso}, \citenamefont {de~Gironcoli}, \citenamefont {Delugas},
  \citenamefont {Jr}, \citenamefont {Ferretti}, \citenamefont {Floris},
  \citenamefont {Fratesi}, \citenamefont {Fugallo}, \citenamefont {Gebauer},
  \citenamefont {Gerstmann}, \citenamefont {Giustino}, \citenamefont {Gorni},
  \citenamefont {Jia}, \citenamefont {Kawamura}, \citenamefont {Ko},
  \citenamefont {Kokalj}, \citenamefont {Kucukbenli}, \citenamefont {Lazzeri},
  \citenamefont {Marsili}, \citenamefont {Marzari}, \citenamefont {Mauri},
  \citenamefont {Nguyen}, \citenamefont {Nguyen}, \citenamefont {de-la Roza},
  \citenamefont {Paulatto}, \citenamefont {Ponce}, \citenamefont {Rocca},
  \citenamefont {Sabatini}, \citenamefont {Santra}, \citenamefont {Schlipf},
  \citenamefont {Seitsonen}, \citenamefont {Smogunov}, \citenamefont {Timrov},
  \citenamefont {Thonhauser}, \citenamefont {Umari}, \citenamefont {Vast},
  \citenamefont {Wu},\ and\ \citenamefont {Baroni}}]{QE-2017}%
  \BibitemOpen
  \bibfield  {author} {\bibinfo {author} {\bibfnamefont {P}~\bibnamefont
  {Giannozzi}}, \bibinfo {author} {\bibfnamefont {O}~\bibnamefont {Andreussi}},
  \bibinfo {author} {\bibfnamefont {T}~\bibnamefont {Brumme}}, \bibinfo
  {author} {\bibfnamefont {O}~\bibnamefont {Bunau}}, \bibinfo {author}
  {\bibfnamefont {M~Buongiorno}\ \bibnamefont {Nardelli}}, \bibinfo {author}
  {\bibfnamefont {M}~\bibnamefont {Calandra}}, \bibinfo {author} {\bibfnamefont
  {R}~\bibnamefont {Car}}, \bibinfo {author} {\bibfnamefont {C}~\bibnamefont
  {Cavazzoni}}, \bibinfo {author} {\bibfnamefont {D}~\bibnamefont {Ceresoli}},
  \bibinfo {author} {\bibfnamefont {M}~\bibnamefont {Cococcioni}}, \bibinfo
  {author} {\bibfnamefont {N}~\bibnamefont {Colonna}}, \bibinfo {author}
  {\bibfnamefont {I}~\bibnamefont {Carnimeo}}, \bibinfo {author} {\bibfnamefont
  {A~Dal}\ \bibnamefont {Corso}}, \bibinfo {author} {\bibfnamefont
  {S}~\bibnamefont {de~Gironcoli}}, \bibinfo {author} {\bibfnamefont
  {P}~\bibnamefont {Delugas}}, \bibinfo {author} {\bibfnamefont {R~A~DiStasio}\
  \bibnamefont {Jr}}, \bibinfo {author} {\bibfnamefont {A}~\bibnamefont
  {Ferretti}}, \bibinfo {author} {\bibfnamefont {A}~\bibnamefont {Floris}},
  \bibinfo {author} {\bibfnamefont {G}~\bibnamefont {Fratesi}}, \bibinfo
  {author} {\bibfnamefont {G}~\bibnamefont {Fugallo}}, \bibinfo {author}
  {\bibfnamefont {R}~\bibnamefont {Gebauer}}, \bibinfo {author} {\bibfnamefont
  {U}~\bibnamefont {Gerstmann}}, \bibinfo {author} {\bibfnamefont
  {F}~\bibnamefont {Giustino}}, \bibinfo {author} {\bibfnamefont
  {T}~\bibnamefont {Gorni}}, \bibinfo {author} {\bibfnamefont {J}~\bibnamefont
  {Jia}}, \bibinfo {author} {\bibfnamefont {M}~\bibnamefont {Kawamura}},
  \bibinfo {author} {\bibfnamefont {H-Y}\ \bibnamefont {Ko}}, \bibinfo {author}
  {\bibfnamefont {A}~\bibnamefont {Kokalj}}, \bibinfo {author} {\bibfnamefont
  {E}~\bibnamefont {Kucukbenli}}, \bibinfo {author} {\bibfnamefont
  {M}~\bibnamefont {Lazzeri}}, \bibinfo {author} {\bibfnamefont
  {M}~\bibnamefont {Marsili}}, \bibinfo {author} {\bibfnamefont
  {N}~\bibnamefont {Marzari}}, \bibinfo {author} {\bibfnamefont
  {F}~\bibnamefont {Mauri}}, \bibinfo {author} {\bibfnamefont {N~L}\
  \bibnamefont {Nguyen}}, \bibinfo {author} {\bibfnamefont {H-V}\ \bibnamefont
  {Nguyen}}, \bibinfo {author} {\bibfnamefont {A~Otero}\ \bibnamefont {de-la
  Roza}}, \bibinfo {author} {\bibfnamefont {L}~\bibnamefont {Paulatto}},
  \bibinfo {author} {\bibfnamefont {S}~\bibnamefont {Ponce}}, \bibinfo {author}
  {\bibfnamefont {D}~\bibnamefont {Rocca}}, \bibinfo {author} {\bibfnamefont
  {R}~\bibnamefont {Sabatini}}, \bibinfo {author} {\bibfnamefont
  {B}~\bibnamefont {Santra}}, \bibinfo {author} {\bibfnamefont {M}~\bibnamefont
  {Schlipf}}, \bibinfo {author} {\bibfnamefont {A~P}\ \bibnamefont
  {Seitsonen}}, \bibinfo {author} {\bibfnamefont {A}~\bibnamefont {Smogunov}},
  \bibinfo {author} {\bibfnamefont {I}~\bibnamefont {Timrov}}, \bibinfo
  {author} {\bibfnamefont {T}~\bibnamefont {Thonhauser}}, \bibinfo {author}
  {\bibfnamefont {P}~\bibnamefont {Umari}}, \bibinfo {author} {\bibfnamefont
  {N}~\bibnamefont {Vast}}, \bibinfo {author} {\bibfnamefont {X}~\bibnamefont
  {Wu}}, \ and\ \bibinfo {author} {\bibfnamefont {S}~\bibnamefont {Baroni}},\
  }\bibfield  {title} {\enquote {\bibinfo {title} {{Advanced capabilities for
  materials modelling with QUANTUM ESPRESSO}},}\ }\href
  {http://stacks.iop.org/0953-8984/29/i=46/a=465901} {\bibfield  {journal}
  {\bibinfo  {journal} {Journal of Physics: Condensed Matter}\ }\textbf
  {\bibinfo {volume} {29}},\ \bibinfo {pages} {465901} (\bibinfo {year}
  {2017})}\BibitemShut {NoStop}%
\bibitem [{\citenamefont {{Ca.pbe-spn-rrkjus\_psl.0.2.3.UPF}}\ \emph
  {et~al.}()\citenamefont {{Ca.pbe-spn-rrkjus\_psl.0.2.3.UPF}}, \citenamefont
  {{Bi.pbe-dn-rrkjus\_psl.0.2.2.UPF}},\ and\ \citenamefont
  {{Bi.rel-pbe-dn-rrkjus\_psl.0.2.2.UPF}}}]{pseudo}%
  \BibitemOpen
  \bibfield  {author} {\bibinfo {author} {\bibfnamefont {{The following
  pseudopotentials were used:}}\ \bibnamefont
  {{Ca.pbe-spn-rrkjus\_psl.0.2.3.UPF}}}, \bibinfo {author} {\bibnamefont
  {{Bi.pbe-dn-rrkjus\_psl.0.2.2.UPF}}}, \ and\ \bibinfo {author} {\bibnamefont
  {{Bi.rel-pbe-dn-rrkjus\_psl.0.2.2.UPF}}},\ }\href@noop {} {\bibinfo
  {journal} {http://www.quantum-espresso.org/pseudopotentials/}\ }\BibitemShut
  {NoStop}%
\bibitem [{\citenamefont {Perdew}\ \emph {et~al.}(1996)\citenamefont {Perdew},
  \citenamefont {Burke},\ and\ \citenamefont {Ernzerhof}}]{pbe}%
  \BibitemOpen
\bibfield  {journal} {  }\bibfield  {author} {\bibinfo {author} {\bibfnamefont
  {John~P.}\ \bibnamefont {Perdew}}, \bibinfo {author} {\bibfnamefont {Kieron}\
  \bibnamefont {Burke}}, \ and\ \bibinfo {author} {\bibfnamefont {Matthias}\
  \bibnamefont {Ernzerhof}},\ }\bibfield  {title} {\enquote {\bibinfo {title}
  {{Generalized Gradient Approximation Made Simple}},}\ }\href {\doibase
  10.1103/PhysRevLett.77.3865} {\bibfield  {journal} {\bibinfo  {journal}
  {Phys. Rev. Lett.}\ }\textbf {\bibinfo {volume} {77}},\ \bibinfo {pages}
  {3865--3868} (\bibinfo {year} {1996})}\BibitemShut {NoStop}%
\bibitem [{\citenamefont {Baroni}\ \emph {et~al.}(2001)\citenamefont {Baroni},
  \citenamefont {de~Gironcoli}, \citenamefont {Dal~Corso},\ and\ \citenamefont
  {Giannozzi}}]{dfpt}%
  \BibitemOpen
  \bibfield  {author} {\bibinfo {author} {\bibfnamefont {Stefano}\ \bibnamefont
  {Baroni}}, \bibinfo {author} {\bibfnamefont {Stefano}\ \bibnamefont
  {de~Gironcoli}}, \bibinfo {author} {\bibfnamefont {Andrea}\ \bibnamefont
  {Dal~Corso}}, \ and\ \bibinfo {author} {\bibfnamefont {Paolo}\ \bibnamefont
  {Giannozzi}},\ }\bibfield  {title} {\enquote {\bibinfo {title} {{Phonons and
  related crystal properties from density-functional perturbation theory}},}\
  }\href {\doibase 10.1103/RevModPhys.73.515} {\bibfield  {journal} {\bibinfo
  {journal} {Rev. Mod. Phys.}\ }\textbf {\bibinfo {volume} {73}},\ \bibinfo
  {pages} {515--562} (\bibinfo {year} {2001})}\BibitemShut {NoStop}%
\bibitem [{\citenamefont {Allen}\ and\ \citenamefont
  {Dynes}(1975)}]{allen-dynes}%
  \BibitemOpen
  \bibfield  {author} {\bibinfo {author} {\bibfnamefont {P.~B.}\ \bibnamefont
  {Allen}}\ and\ \bibinfo {author} {\bibfnamefont {R.~C.}\ \bibnamefont
  {Dynes}},\ }\bibfield  {title} {\enquote {\bibinfo {title} {{Transition
  temperature of strong-coupled superconductors reanalyzed}},}\ }\href
  {\doibase 10.1103/PhysRevB.12.905} {\bibfield  {journal} {\bibinfo  {journal}
  {Phys. Rev. B}\ }\textbf {\bibinfo {volume} {12}},\ \bibinfo {pages}
  {905--922} (\bibinfo {year} {1975})}\BibitemShut {NoStop}%
\bibitem [{\citenamefont {Madsen}\ and\ \citenamefont
  {Singh}(2006)}]{boltztrap}%
  \BibitemOpen
  \bibfield  {author} {\bibinfo {author} {\bibfnamefont {Georg~K.H.}\
  \bibnamefont {Madsen}}\ and\ \bibinfo {author} {\bibfnamefont {David~J.}\
  \bibnamefont {Singh}},\ }\bibfield  {title} {\enquote {\bibinfo {title}
  {{{BoltzTraP. A code for calculating band-structure dependent
  quantities}}},}\ }\href {\doibase 10.1016/j.cpc.2006.03.007} {\bibfield
  {journal} {\bibinfo  {journal} {Computer Physics Communications}\ }\textbf
  {\bibinfo {volume} {175}},\ \bibinfo {pages} {67--71} (\bibinfo {year}
  {2006})}\BibitemShut {NoStop}%
\bibitem [{\citenamefont {Grimvall}(1981)}]{grimvall}%
  \BibitemOpen
  \bibfield  {author} {\bibinfo {author} {\bibfnamefont {G.}~\bibnamefont
  {Grimvall}},\ }\href@noop {} {\emph {\bibinfo {title} {The electron-phonon
  interaction in metals}}}\ (\bibinfo  {publisher} {North-Holland, Amsterdam},\
  \bibinfo {year} {1981})\BibitemShut {NoStop}%
\bibitem [{\citenamefont {Stewart}(2008)}]{palladium}%
  \BibitemOpen
  \bibfield  {author} {\bibinfo {author} {\bibfnamefont {Derek~A}\ \bibnamefont
  {Stewart}},\ }\bibfield  {title} {\enquote {\bibinfo {title} {{Ab initio
  investigation of phonon dispersion and anomalies in palladium}},}\ }\href
  {http://stacks.iop.org/1367-2630/10/i=4/a=043025} {\bibfield  {journal}
  {\bibinfo  {journal} {New Journal of Physics}\ }\textbf {\bibinfo {volume}
  {10}},\ \bibinfo {pages} {043025} (\bibinfo {year} {2008})}\BibitemShut
  {NoStop}%
\bibitem [{\citenamefont {{Kohn}}(1959)}]{kohn-anomaly}%
  \BibitemOpen
  \bibfield  {author} {\bibinfo {author} {\bibfnamefont {W.}~\bibnamefont
  {{Kohn}}},\ }\bibfield  {title} {\enquote {\bibinfo {title} {{Image of the
  {F}ermi Surface in the Vibration Spectrum of a Metal}},}\ }\href {\doibase
  10.1103/PhysRevLett.2.393} {\bibfield  {journal} {\bibinfo  {journal}
  {Physical Review Letters}\ }\textbf {\bibinfo {volume} {2}},\ \bibinfo
  {pages} {393--394} (\bibinfo {year} {1959})}\BibitemShut {NoStop}%
\bibitem [{\citenamefont {Wierzbowska}\ \emph {et~al.}(2005)\citenamefont
  {Wierzbowska}, \citenamefont {de~Gironcoli},\ and\ \citenamefont
  {Giannozzi}}]{wierzbowska}%
  \BibitemOpen
  \bibfield  {author} {\bibinfo {author} {\bibfnamefont {Malgorzata}\
  \bibnamefont {Wierzbowska}}, \bibinfo {author} {\bibfnamefont {Stefano}\
  \bibnamefont {de~Gironcoli}}, \ and\ \bibinfo {author} {\bibfnamefont
  {Paolo}\ \bibnamefont {Giannozzi}},\ }\bibfield  {title} {\enquote {\bibinfo
  {title} {{Origins of low-and high-pressure discontinuities of $T_c$ in
  niobium}},}\ }\href@noop {} {\bibfield  {journal} {\bibinfo  {journal} {arXiv
  preprint cond-mat/0504077}\ } (\bibinfo {year} {2005})}\BibitemShut {NoStop}%
\bibitem [{Note2()}]{Note2}%
  \BibitemOpen
  \bibinfo {note} {See Fig. S2 in Supplemental Material for the phonon
  displacement patterns in A point.}\BibitemShut {Stop}%
\bibitem [{\citenamefont {McMillan}(1968)}]{mcmillan}%
  \BibitemOpen
  \bibfield  {author} {\bibinfo {author} {\bibfnamefont {W.~L.}\ \bibnamefont
  {McMillan}},\ }\bibfield  {title} {\enquote {\bibinfo {title} {Transition
  temperature of strong-coupled superconductors},}\ }\href {\doibase
  10.1103/PhysRev.167.331} {\bibfield  {journal} {\bibinfo  {journal} {Phys.
  Rev.}\ }\textbf {\bibinfo {volume} {167}},\ \bibinfo {pages} {331--344}
  (\bibinfo {year} {1968})}\BibitemShut {NoStop}%
\bibitem [{Note3()}]{Note3}%
  \BibitemOpen
  \bibinfo {note} {It is worth recalling here, that McMillan~\cite {mcmillan}
  for his $T_c$ formula assumed $\mu ^* = 0.13$ for transition metals and $\mu
  ^* = 0.10$ for simple metals, whereas Allen and Dynes~\cite {allen-dynes}
  recommended using their formula with $\mu ^* = 0.10$ for transition metals,
  and even lower values, like $\mu ^* = 0.09$, for simple metals. We
  consistently use $\mu ^* = 0.10$ here. Taking $\mu ^* = 0.13$ with McMillan
  formula and experimental $T_c$ for CaBi$_2$ gives slightly larger $\lambda
  _{\protect \rm expt} = 0.59$.}\BibitemShut {Stop}%
\bibitem [{Note4()}]{Note4}%
  \BibitemOpen
  \bibinfo {note} {See Fig. S3 in Supplemental Material for the phonon
  displacement patterns in Y point.}\BibitemShut {Stop}%
\bibitem [{\citenamefont {Chen}(2018)}]{kbi2-elph}%
  \BibitemOpen
  \bibfield  {author} {\bibinfo {author} {\bibfnamefont {Jianyong}\
  \bibnamefont {Chen}},\ }\bibfield  {title} {\enquote {\bibinfo {title} {{A
  Comprehensive Investigation of Superconductor KBi 2 via First-Principles
  Calculations}},}\ }\href@noop {} {\bibfield  {journal} {\bibinfo  {journal}
  {Journal of Superconductivity and Novel Magnetism}\ }\textbf {\bibinfo
  {volume} {31}},\ \bibinfo {pages} {1301--1307} (\bibinfo {year}
  {2018})}\BibitemShut {NoStop}%
\bibitem [{\citenamefont {Sun}\ \emph {et~al.}(2016)\citenamefont {Sun},
  \citenamefont {Liu},\ and\ \citenamefont {Lei}}]{kbi2-exp}%
  \BibitemOpen
  \bibfield  {author} {\bibinfo {author} {\bibfnamefont {Shanshan}\
  \bibnamefont {Sun}}, \bibinfo {author} {\bibfnamefont {Kai}\ \bibnamefont
  {Liu}}, \ and\ \bibinfo {author} {\bibfnamefont {Hechang}\ \bibnamefont
  {Lei}},\ }\bibfield  {title} {\enquote {\bibinfo {title} {{Type-I
  superconductivity in KBi2 single crystals}},}\ }\href@noop {} {\bibfield
  {journal} {\bibinfo  {journal} {Journal of Physics: Condensed Matter}\
  }\textbf {\bibinfo {volume} {28}},\ \bibinfo {pages} {085701} (\bibinfo
  {year} {2016})}\BibitemShut {NoStop}%
\bibitem [{\citenamefont {Jha}\ \emph {et~al.}(2016)\citenamefont {Jha},
  \citenamefont {Avila},\ and\ \citenamefont {Ribeiro}}]{abi3-b}%
  \BibitemOpen
  \bibfield  {author} {\bibinfo {author} {\bibfnamefont {Rajveer}\ \bibnamefont
  {Jha}}, \bibinfo {author} {\bibfnamefont {Marcos~A}\ \bibnamefont {Avila}}, \
  and\ \bibinfo {author} {\bibfnamefont {Raquel~A}\ \bibnamefont {Ribeiro}},\
  }\bibfield  {title} {\enquote {\bibinfo {title} {{Hydrostatic pressure effect
  on the superconducting properties of BaBi${}_3$ and SrBi${}_3$ single
  crystals}},}\ }\href@noop {} {\bibfield  {journal} {\bibinfo  {journal}
  {Superconductor Science and Technology}\ }\textbf {\bibinfo {volume} {30}},\
  \bibinfo {pages} {025015} (\bibinfo {year} {2016})}\BibitemShut {NoStop}%
\bibitem [{\citenamefont {Haldolaarachchige}\ \emph {et~al.}(2014)\citenamefont
  {Haldolaarachchige}, \citenamefont {Kushwaha}, \citenamefont {Gibson},\ and\
  \citenamefont {Cava}}]{babi3-h}%
  \BibitemOpen
  \bibfield  {author} {\bibinfo {author} {\bibfnamefont {Neel}\ \bibnamefont
  {Haldolaarachchige}}, \bibinfo {author} {\bibfnamefont {SK}~\bibnamefont
  {Kushwaha}}, \bibinfo {author} {\bibfnamefont {Quinn}\ \bibnamefont
  {Gibson}}, \ and\ \bibinfo {author} {\bibfnamefont {RJ}~\bibnamefont
  {Cava}},\ }\bibfield  {title} {\enquote {\bibinfo {title} {{Superconducting
  properties of BaBi${}_3$}},}\ }\href@noop {} {\bibfield  {journal} {\bibinfo
  {journal} {Superconductor Science and Technology}\ }\textbf {\bibinfo
  {volume} {27}},\ \bibinfo {pages} {105001} (\bibinfo {year}
  {2014})}\BibitemShut {NoStop}%
\bibitem [{\citenamefont {Wang}\ \emph {et~al.}(2018)\citenamefont {Wang},
  \citenamefont {Luo}, \citenamefont {Ishigaki}, \citenamefont {Matsubayashi},
  \citenamefont {Cheng}, \citenamefont {Sun},\ and\ \citenamefont
  {Uwatoko}}]{abi3-exp-j}%
  \BibitemOpen
  \bibfield  {author} {\bibinfo {author} {\bibfnamefont {Bosen}\ \bibnamefont
  {Wang}}, \bibinfo {author} {\bibfnamefont {Xuan}\ \bibnamefont {Luo}},
  \bibinfo {author} {\bibfnamefont {Kento}\ \bibnamefont {Ishigaki}}, \bibinfo
  {author} {\bibfnamefont {Kazuyuki}\ \bibnamefont {Matsubayashi}}, \bibinfo
  {author} {\bibfnamefont {Jinguang}\ \bibnamefont {Cheng}}, \bibinfo {author}
  {\bibfnamefont {Yuping}\ \bibnamefont {Sun}}, \ and\ \bibinfo {author}
  {\bibfnamefont {Yoshiya}\ \bibnamefont {Uwatoko}},\ }\bibfield  {title}
  {\enquote {\bibinfo {title} {{Two distinct superconducting phases and
  pressure-induced crossover from type-II to type-I superconductivity in the
  spin-orbit-coupled superconductors $\mathrm{BaB}{\mathrm{i}}_{3}$ and
  $\mathrm{SrB}{\mathrm{i}}_{3}$}},}\ }\href {\doibase
  10.1103/PhysRevB.98.220506} {\bibfield  {journal} {\bibinfo  {journal} {Phys.
  Rev. B}\ }\textbf {\bibinfo {volume} {98}},\ \bibinfo {pages} {220506}
  (\bibinfo {year} {2018})}\BibitemShut {NoStop}%
\bibitem [{\citenamefont {Kakihana}\ \emph {et~al.}(2015)\citenamefont
  {Kakihana}, \citenamefont {Akamine}, \citenamefont {Yara}, \citenamefont
  {Teruya}, \citenamefont {Nakamura}, \citenamefont {Takeuchi}, \citenamefont
  {Hedo}, \citenamefont {Nakama}, \citenamefont {{\=O}nuki},\ and\
  \citenamefont {Harima}}]{srbi3-k}%
  \BibitemOpen
  \bibfield  {author} {\bibinfo {author} {\bibfnamefont {Masashi}\ \bibnamefont
  {Kakihana}}, \bibinfo {author} {\bibfnamefont {Hiromu}\ \bibnamefont
  {Akamine}}, \bibinfo {author} {\bibfnamefont {Tomoyuki}\ \bibnamefont
  {Yara}}, \bibinfo {author} {\bibfnamefont {Atsushi}\ \bibnamefont {Teruya}},
  \bibinfo {author} {\bibfnamefont {Ai}~\bibnamefont {Nakamura}}, \bibinfo
  {author} {\bibfnamefont {Tetsuya}\ \bibnamefont {Takeuchi}}, \bibinfo
  {author} {\bibfnamefont {Masato}\ \bibnamefont {Hedo}}, \bibinfo {author}
  {\bibfnamefont {Takao}\ \bibnamefont {Nakama}}, \bibinfo {author}
  {\bibfnamefont {Yoshichika}\ \bibnamefont {{\=O}nuki}}, \ and\ \bibinfo
  {author} {\bibfnamefont {Hisatomo}\ \bibnamefont {Harima}},\ }\bibfield
  {title} {\enquote {\bibinfo {title} {{Fermi Surface Properties Based on the
  Relativistic Effect in SrBi$_3$ with AuCu$_3$-Type Cubic Structure}},}\
  }\href@noop {} {\bibfield  {journal} {\bibinfo  {journal} {Journal of the
  Physical Society of Japan}\ }\textbf {\bibinfo {volume} {84}},\ \bibinfo
  {pages} {124702} (\bibinfo {year} {2015})}\BibitemShut {NoStop}%
\bibitem [{\citenamefont {T\"ut\"unc\"u}\ \emph {et~al.}(2018)\citenamefont
  {T\"ut\"unc\"u}, \citenamefont {Karaca}, \citenamefont {Uzunok},\ and\
  \citenamefont {Srivastava}}]{labi3-elph}%
  \BibitemOpen
  \bibfield  {author} {\bibinfo {author} {\bibfnamefont {H.~M.}\ \bibnamefont
  {T\"ut\"unc\"u}}, \bibinfo {author} {\bibfnamefont
  {Ertu\ifmmode\check{g}\else\v{g}\fi{}rul}\ \bibnamefont {Karaca}}, \bibinfo
  {author} {\bibfnamefont {H.~Y.}\ \bibnamefont {Uzunok}}, \ and\ \bibinfo
  {author} {\bibfnamefont {G.~P.}\ \bibnamefont {Srivastava}},\ }\bibfield
  {title} {\enquote {\bibinfo {title} {{Role of spin-orbit coupling in the
  physical properties of $\mathrm{La}{X}_{3}$ ($X=\mathrm{In}$, P, Bi)
  superconductors}},}\ }\href {\doibase 10.1103/PhysRevB.97.174512} {\bibfield
  {journal} {\bibinfo  {journal} {Phys. Rev. B}\ }\textbf {\bibinfo {volume}
  {97}},\ \bibinfo {pages} {174512} (\bibinfo {year} {2018})}\BibitemShut
  {NoStop}%
\bibitem [{\citenamefont {Dong}\ and\ \citenamefont {Fan}(2015)}]{cabi3}%
  \BibitemOpen
  \bibfield  {author} {\bibinfo {author} {\bibfnamefont {Xu}~\bibnamefont
  {Dong}}\ and\ \bibinfo {author} {\bibfnamefont {Changzeng}\ \bibnamefont
  {Fan}},\ }\bibfield  {title} {\enquote {\bibinfo {title} {{Rich
  stoichiometries of stable Ca-Bi system: Structure prediction and
  superconductivity}},}\ }\href@noop {} {\bibfield  {journal} {\bibinfo
  {journal} {Scientific reports}\ }\textbf {\bibinfo {volume} {5}},\ \bibinfo
  {pages} {9326} (\bibinfo {year} {2015})}\BibitemShut {NoStop}%
\bibitem [{\citenamefont {Fujimori}\ \emph {et~al.}(2000)\citenamefont
  {Fujimori}, \citenamefont {Kan}, \citenamefont {Shinozaki},\ and\
  \citenamefont {Kawaguti}}]{nibi3-j}%
  \BibitemOpen
  \bibfield  {author} {\bibinfo {author} {\bibfnamefont {Yasunobu}\
  \bibnamefont {Fujimori}}, \bibinfo {author} {\bibfnamefont {Shin-ichi}\
  \bibnamefont {Kan}}, \bibinfo {author} {\bibfnamefont {Bunjyu}\ \bibnamefont
  {Shinozaki}}, \ and\ \bibinfo {author} {\bibfnamefont {Takasi}\ \bibnamefont
  {Kawaguti}},\ }\bibfield  {title} {\enquote {\bibinfo {title}
  {{Superconducting and normal state properties of NiBi${}_3$}},}\ }\href@noop
  {} {\bibfield  {journal} {\bibinfo  {journal} {Journal of the Physical
  Society of Japan}\ }\textbf {\bibinfo {volume} {69}},\ \bibinfo {pages}
  {3017--3026} (\bibinfo {year} {2000})}\BibitemShut {NoStop}%
\bibitem [{\citenamefont {Kumar}\ \emph {et~al.}(2011)\citenamefont {Kumar},
  \citenamefont {Kumar}, \citenamefont {Vajpayee}, \citenamefont {Gahtori},
  \citenamefont {Sharma}, \citenamefont {Ahluwalia}, \citenamefont {Auluck},\
  and\ \citenamefont {Awana}}]{nibi3-i}%
  \BibitemOpen
  \bibfield  {author} {\bibinfo {author} {\bibfnamefont {Jagdish}\ \bibnamefont
  {Kumar}}, \bibinfo {author} {\bibfnamefont {Anuj}\ \bibnamefont {Kumar}},
  \bibinfo {author} {\bibfnamefont {Arpita}\ \bibnamefont {Vajpayee}}, \bibinfo
  {author} {\bibfnamefont {Bhasker}\ \bibnamefont {Gahtori}}, \bibinfo {author}
  {\bibfnamefont {Devina}\ \bibnamefont {Sharma}}, \bibinfo {author}
  {\bibfnamefont {PK}~\bibnamefont {Ahluwalia}}, \bibinfo {author}
  {\bibfnamefont {S}~\bibnamefont {Auluck}}, \ and\ \bibinfo {author}
  {\bibfnamefont {VPS}\ \bibnamefont {Awana}},\ }\bibfield  {title} {\enquote
  {\bibinfo {title} {{Physical property and electronic structure
  characterization of bulk superconducting Bi${}_3$Ni}},}\ }\href@noop {}
  {\bibfield  {journal} {\bibinfo  {journal} {Superconductor Science and
  Technology}\ }\textbf {\bibinfo {volume} {24}},\ \bibinfo {pages} {085002}
  (\bibinfo {year} {2011})}\BibitemShut {NoStop}%
\bibitem [{Note5()}]{Note5}%
  \BibitemOpen
  \bibinfo {note} {Sommerfeld coefficient $\gamma _{\protect \rm expt}$ may be
  renormalized by other effects than the electron-phonon interaction, however
  in such intermetallic compounds, with $s$ and $p$ electrons at the Fermi
  level, strong electron correlactions or paramagnons are not expected to
  appear.}\BibitemShut {Stop}%
\end{thebibliography}%

\newpage

\section{Supplemental Material}

Fig. S1 shows in a convenient way relation between the primitive and conventional base-centered crystal cells of CaBi$_2$, which help to analyze Figs. S2 and S3.

Fig. S2 show phonon displacement patterns of all 18 phonon modes in A-point from the scalar-relativistic calculations, as in this case phonon linewidths are very large. Amplitude of modes is enlarged.

Fig. S3 show phonon displacement patterns of all 18 phonon modes in Y-point from the relativistic calculations. This point was chosen as it has large phonon linewidths in the relativistic case. Amplitude of modes is enlarged.

\begin{figure}[htb!]
\includegraphics[width=0.70\textwidth]{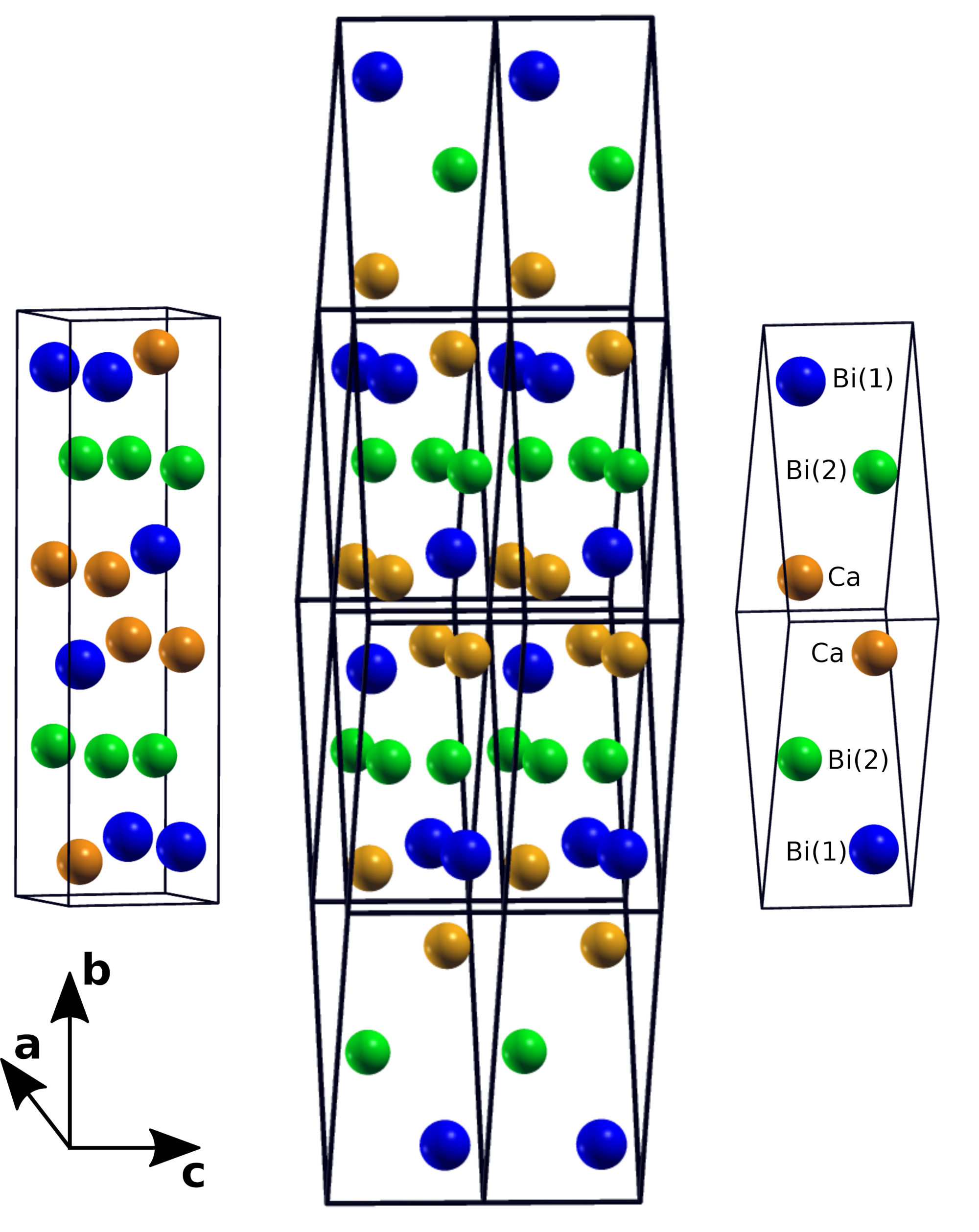}
\caption{Crystal structure of CaBi$_2$. Left: Conventional unit cell, with Bi(1)-Ca and Bi(2) layers, perpendicular to {\it b} axis. Right: primitive cell. Middle: 2x2x2 primitive supercell, which shows the relation of primitive and conventional unit cells. {\it a,c} are the in-plane directions, whereas {\it b} is perpendicular to Bi and Ca-Bi layers.}
\end{figure}

\begin{figure}[htb]
\includegraphics[width=0.9\textwidth]{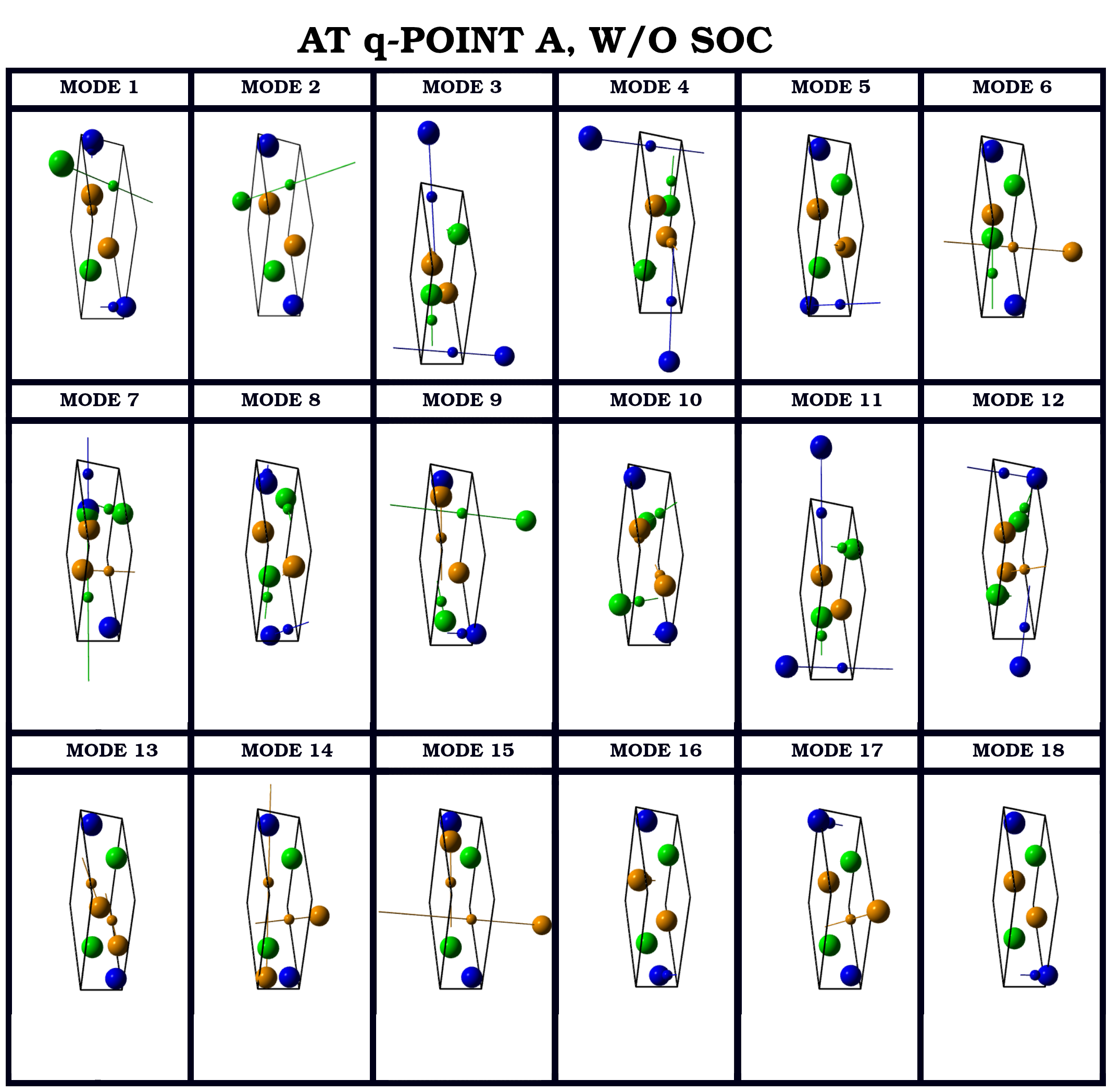}
\caption{Phonon displacement patterns of all 18 phonon modes in A-point from the scalar-relativistic calculations. Fig. S1 helps to identify which directions are in-plane or perpendicular to atomic planes.}
\end{figure}

\begin{figure}[htb]
\includegraphics[width=0.9\textwidth]{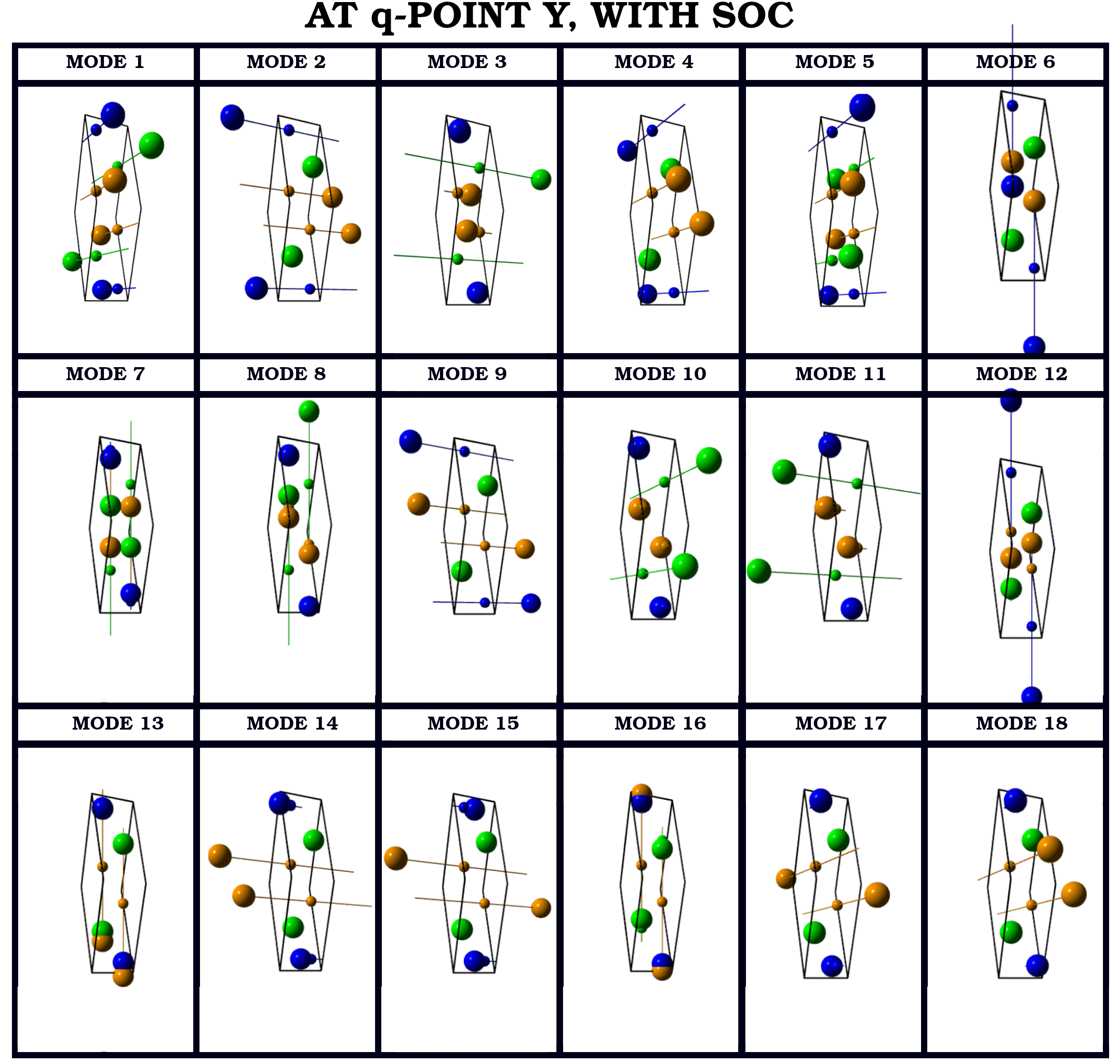}
\caption{Phonon displacement patterns of all 18 phonon modes in Y-point from the relativistic calculations. Fig. S1 helps to identify which directions are in-plane or perpendicular to atomic planes.}
\end{figure}

\end{document}